\documentclass[%
 reprint,
superscriptaddress,
 amsmath,amssymb,
 prl,
]{revtex4-2}

\usepackage[normalem]{ulem}
\usepackage{graphicx}
\usepackage{dcolumn}
\usepackage{bm}

\usepackage[utf8]{inputenc}
\usepackage[T1]{fontenc}
\usepackage{mathptmx}
\usepackage{etoolbox}
\usepackage{verbatim}
\usepackage{float}
\usepackage{caption}
\usepackage{subcaption}
\usepackage{amsmath}
\usepackage{amssymb}
\usepackage{xcolor}
\usepackage{hyperref}
\newcommand{\CR}[1]{{\color{black} #1}}
\newcommand{\sm}[1]{{\color{black} #1}}
\newcommand{\AC}[1]{{\color{black} #1}}

\begin{document}


\title{Intrinsic femtosecond structure of extreme contrast harmonic pulses:\\ influence on relativistic laser-solid interactions}

\author{C. Aparajit}
 \affiliation{
Tata Institute of Fundamental Research, 1 Homi Bhabha Road, Colaba, Mumbai 400 005, India
}

\author{Anandam Choudhary}
\affiliation{
Tata Institute of Fundamental Research, 1 Homi Bhabha Road, Colaba, Mumbai 400 005, India
}
\author{Ankit Dulat}
\affiliation{
Tata Institute of Fundamental Research, 1 Homi Bhabha Road, Colaba, Mumbai 400 005, India
}

\author{Mickael Grech}
\affiliation{LULI, CNRS, CEA, Sorbonne Université, École Polytechnique, F-91120 Palaiseau, France}

\author{Samuel Marini}
\affiliation{CEA, Université Paris-Saclay, IRFU, DACM, Gif-sur-Yvette, France}

\author{Amit D. Lad}
\affiliation{
Tata Institute of Fundamental Research, 1 Homi Bhabha Road, Colaba, Mumbai 400 005, India
}

\author{Yash M. Ved}
\affiliation{
Tata Institute of Fundamental Research, 1 Homi Bhabha Road, Colaba, Mumbai 400 005, India
}

\author{Michèle Raynaud}
\affiliation{
LSI, CEA, CNRS, École Polytechnique, Institut Polytechnique,  F-91120 Palaiseau, France}

\author{Caterina Riconda}
\affiliation{LULI, CNRS, CEA, Sorbonne Université, École Polytechnique,  F-91120 Palaiseau, France}

\author{G. Ravindra Kumar}
\affiliation{
Tata Institute of Fundamental Research, 1 Homi Bhabha Road, Colaba, Mumbai 400 005, India }
\email{grk@tifr.res.in}

\date{\today}

\begin{abstract}
Extreme intensity contrast is considered essential for ultraintense, femtosecond laser excitation of solid targets, in particular for studies with structured or ultra-thin targets. Second-harmonic generation has been used to maximize the contrast in the nanosecond and picosecond timescales but the resulting pulses can have intense broad femtosecond structures in the rising edge of the pulse.  We show that femtosecond scale structures that arise in this process critically modify the interaction, \AC{by altering the local field structures and hence redirecting the electron trajectories and distributions}, especially concerning resonant phenomena such as surface plasmon excitation in structured targets. Particle-in-cell (PIC) simulations fully support and \CR{give further insight into our experimental results}. 
\CR{Our findings have important implications not only for the use of harmonic pulses on solid targets but also for two-color schemes based on second harmonic pulses}. 
\end{abstract}

\maketitle

Rapid progress in high-power laser and high energy density science has led to a multitude of applications, for example, the generation and control of high brightness, ultrafast electromagnetic and particle sources, laboratory simulation of astrophysical scenarios, application of extreme shocks to materials, etc. \cite{drake_2006,  Kaw2017}. The physics of the interactions driven by such femtosecond terawatt and petawatt laser systems is highly nonlinear and complex. To produce the cleanest interactions of relativistic lasers with solid targets, avoiding any pre-plasma, experiments demand ultrahigh intensity contrast, {\it i.e.}, the intensity on the picosecond and nanosecond scales preceding the femtosecond peak should fall off to a value below the plasma creation threshold for the target. The intensity contrast crucially determines the pre-plasma density gradient and in turn, affects fundamental processes like absorption \cite{Cerchez_PhysRevLett.100.245001, Ping_PhysRevLett.100.085004}, hot electron generation \cite{Kemp_PhysRevLett.101.075004, Peebles_2017}, high-harmonic generation \cite{Kahaly_PhysRevLett.110.175001}, etc. There have been many recent studies demanding ultrahigh contrast pulses, for the generation of homogeneous high-energy-density matter \cite{Beier2022}, high-harmonic generation \cite{Dromey2006, Dollar2013_HHG_Solids_PhysRevLett.110.175002} which crucially depends on the steepness of the plasma density gradient, electron and ion acceleration from ultra-thin targets \cite{Singh2022, Macchi_2010_RPA, Henig_2009_PhysRevLett.103.245003_RPA} and the use of structured targets for creating hot, dense plasma conditions and hence high flux, electromagnetic and material particle emission \cite{Purvis2013, Cantono_PRL2018, Fedeli_PRL_2016, Ceccotti_PRL2013, Rajeev-PRL2003,Rocca_SciAdv, Kahaly-PRL2008, Lad2022}.
\smallskip

An effective approach to enhance the intensity contrast is the use of a second harmonic of the input laser pulse, the harmonic generation process expected to improve the contrast by many orders of magnitude to an "extreme" level \cite{Price_PRL1995,Chen_PhysRevLett.100.045004,Saemann_PRL1999, Purvis2013, Rocca_SciAdv, Hollinger2020, Beier2022}. 
 Moreover, these pulses have crucial implications in two-color relativistic femtosecond laser schemes for applications in particle acceleration, high-harmonic generation, attosecond control, and generation of positron beams, to list a few \cite{Kim_Jong_PRL2005_HHG_twocolor, Yeung_NatPhotonics_2017_attosecond_twocolor, Yue-Yue_PRL2019_positron_twocolor, Song_Li_SciAdv2019_accelerator_twocolor, Barbosa_arvix2023_breitwheeler_twocolor}. 
 However, the impact of the nonlinear process on the intensity profile and instantaneous temporal/spectral phase of the resulting harmonic pulse in the femtosecond regime needs to be known to understand and possibly tailor the interaction with the target. 
Recently, it has been demonstrated that efficient SHG in terawatt/petawatt class laser systems can introduce modulations to the femtosecond temporal pulse profiles. In particular, these can produce femtosecond pre-structures in the laser pulses \cite{Aparajit_APL2023, Aparajit:21}, which can seriously alter the laser-matter interactions, especially with ultrathin targets (< 100 nm), structured targets, nanowire surfaces \cite{Rocca_SciAdv, Purvis2013}, and grating targets \sm{\cite{Cantono_PRL2018, Fedeli_PRL_2016, Ceccotti_PRL2013, Kahaly-PRL2008, Lad2022}. }
\smallskip

Laser pulse shaping has been studied in detail when it comes to nanosecond lasers, especially for applications in inertial confinement fusion (ICF) \cite{Hurricane2014}, x-ray lasers \cite{Cojocaru:16}, laser-plasma-based EUV sources \cite{Nowak:16, Meijer:17}, control of material transport \cite{Pangovski_2014} and more. In the femtosecond and picosecond regimes, pulse shaping is often done in the frequency domain using spatial light modulators or acousto-optic phase dispersive filters, but these are typically done at lower input energies with the goal of studying coherent control and photochemistry \cite{Warren_Science1993, Assion_Science1998}. However, hardly any studies explore the role of resulting temporal structures at the femtosecond scales and at relativistic intensities. 
\smallskip

In parallel studies, laser interactions with grating targets have been explored and key parameters have been identified to efficiently excite surface plasma waves (SPW) to improve laser absorption, and assist in the generation of relativistic hot electrons and steer their directions 
\cite{Riconda_PoP2015_SPW_electron_scaling, Raynaud_PoP2007_SPW, Kupersztych_Raynaud_Riconda_PoP2004_SPW, Raynaud2020, Bigongiari_PoP2013, Marini_PoP2021}. In particular, the generation of surface plasma waves (SPW) is a resonant phenomenon and acts as a sensitive test to study the role of femtosecond scale pre-structures on the fundamental interactions in relativistic laser-solid interactions.
\smallskip

In this Letter, we report the role of femtosecond pre-structures on the excitation of SPW in sub-$\lambda$ grating targets by examining their impact on laser absorption, hot electron angular distributions and energy spectra. We present experimental data using extreme-contrast (<$10^{-12}$), second-harmonic, femtosecond, 400 nm relativistic laser pulse interactions with sub-$\lambda$ gratings as well as flat targets. Experimental results are well supported with 2D particle-in-cell (PIC) simulations performed using the SMILEI code \sm{\cite{Smilei_DEROUILLAT2018351}}.  
\smallskip

The experiment (Fig \ref{fig:setup} (a)) was conducted using a chirped pulse amplified 150 TW Ti: sapphire laser system delivering 800 nm, 27 fs, 50 nm bandwidth pulses at a 5 Hz repetition rate. Before the measurements, the laser was well stabilized (shot-to-shot energy fluctuations $<\pm 5\%$). The 800 nm pulses were up-converted inside a vacuum chamber (10$^{-6}$ torr) to 400 nm using a 70 mm diameter, 2 mm thick type-I lithium triborate (LBO) second-harmonic crystal (Cristal Laser) \cite{Aparajit:21}. A large-aperture half-wave plate was placed before the SHG crystal to ensure that the SH pulses are p-polarized, confirmed using an analyzer/polarizer setup. Multiple dichroic mirrors (at 400 nm) were used in series to remove the residual 800 nm components (800:400 energy ratio was reduced to a level of $\sim$ 10$^{-5}$, measured using pyroelectric detectors (OPHIR PE-50C)). The electron angular distributions were measured using image plate (IP) detectors placed in front of the target to cover most of the front angles (a small hole was made in the IPs to let the laser go through). We used Al-foils to allow only electrons above 100 keV to reach the IP. Three electron spectrometers were also placed to characterize the electron energy spectra in three specific directions, {\it i.e.}, front specular reflection, positive surface (+90$^{\circ}$) and negative surface (+270$^{\circ}$) corresponding to direct laser acceleration and electrons accelerated by the surface modes. Each spectrometer has a 0.1 Tesla magnetic field with an IP as the detector.
The measurable range of energies in these spectrometers is 0.1-7.0 MeV. 

\renewcommand{\thefigure}{1}
\setlength{\belowcaptionskip}{-10pt}
\begin{figure}[!h]
    \centering
    \includegraphics[width = \linewidth]{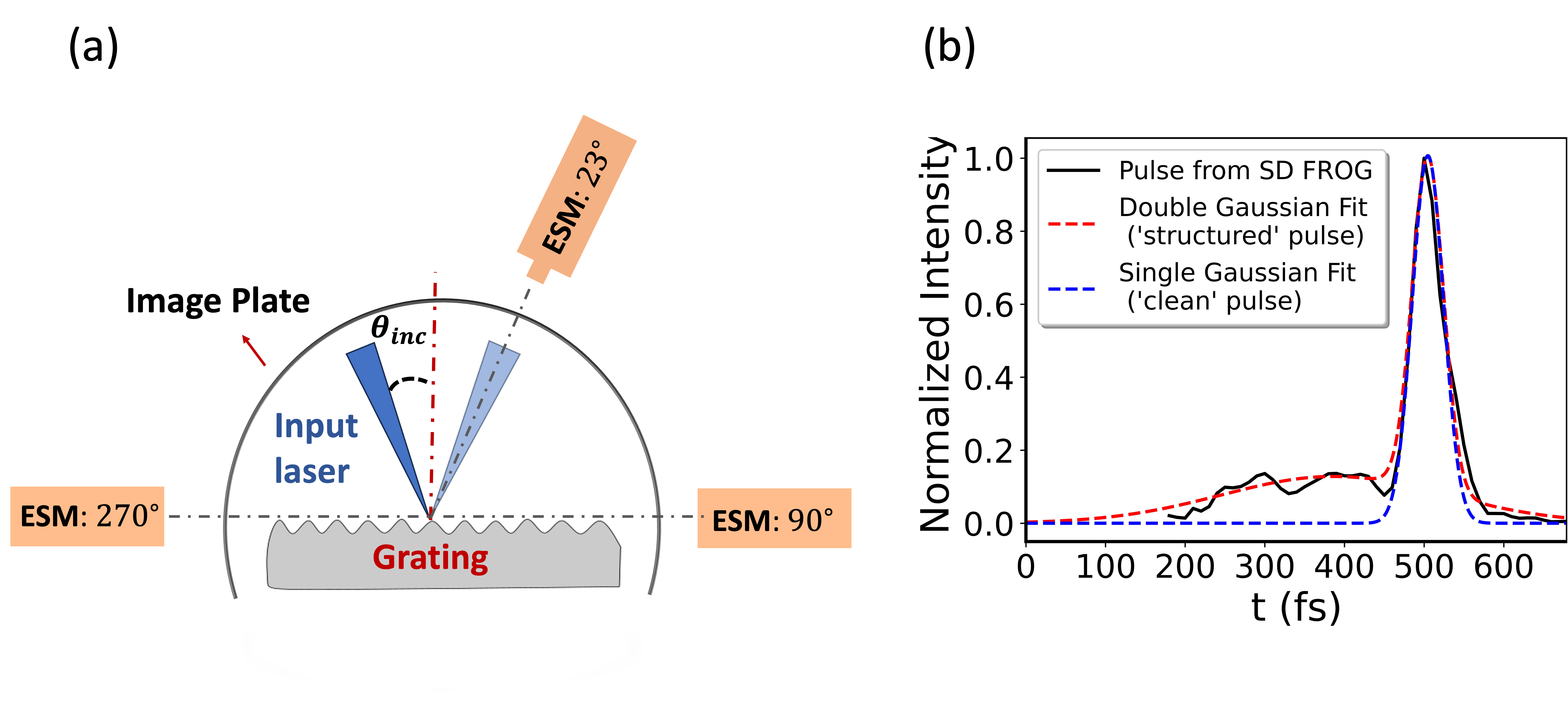}
    \caption{ (a) The technique used for electron angular distributions and hot electron spectra measurements using image plate detectors and electron spectrometer (ESM). (b) Temporal profiles of the input laser pulse (with the femtosecond pre-structure) measured using SD-FROG setup \cite{Aparajit_APL2023}, and the corresponding single and double Gaussian fit temporal profiles ("clean" pulse and "structured" pulse with the femtosecond pre-structure) used in PIC simulations.}
    \label{fig:setup}
\end{figure}

The primary target used in the experiment is a 3600 lines/mm holographic grating target (from Edmund Optics) which is sub-$\lambda$ for our input wavelength of 400 nm. \AC{Two important processes arise in laser interactions with a grating target. Firstly, the structures allow a local enhancement of the electric fields in their vicinity, and in turn, lead to enhanced absorption and hot electron generation \cite{Rajeev_OL2004}. Secondly, they allow resonant excitation of surface plasma waves (SPW) leading to a efficient acceleration and collimation of electrons along the SPW modes. The advantage of a sub-$\lambda$ grating is to reduce the number of diffraction modes and couple the laser energy to only a single resonant surface mode along the negative surface (the surface opposite to the laser direction; 270$^{\circ}$ in  Fig 1(b)}. The sub-$\lambda$ grating used in our experiment shows a resonance for $\theta_{inc}$ (angle of incidence) of 23$^{\circ}$; efficient excitation of surface plasma waves occurs at this angle of incidence. We used a flat BK7 target for comparison. For the electron angular distribution and electron spectra measurements, the peak laser intensity was $8\times 10^{18}$ W/cm$^2$. 
\smallskip

\renewcommand{\thefigure}{2}
\setlength{\belowcaptionskip}{-10pt}
\begin{figure*}[!ht]
    \centering
    \includegraphics[width = 0.85\linewidth]{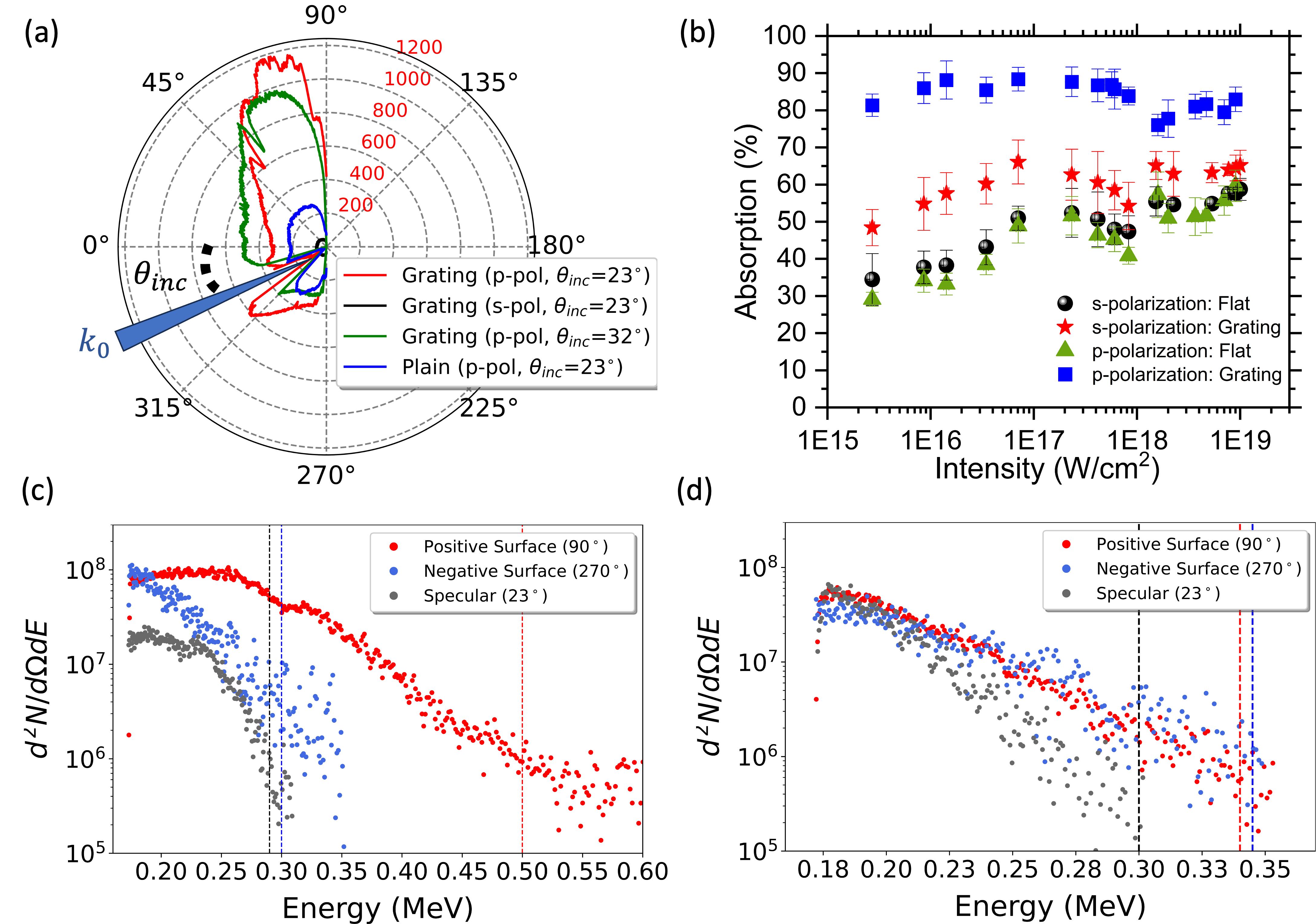}
    \caption{(a) Hot electron angular distribution from experiment for the grating and plain targets. (b) Laser absorption data from experiments for the sub-$\lambda$ grating and plain target for the cases of `p' and `s'-polarized input laser. Experimental hot electron spectra from the (c) grating target and (d) plain target at three different directions, {\it i.e.}, positive surface, negative surface and specular directions. The dashed vertical lines indicate the respective cutoff energies.}
    \label{fig:exp_angdist_esm_combined}
\end{figure*}

The most striking observation pertains to the angular distribution of the ejected fast electrons. (Fig. \ref{fig:exp_angdist_esm_combined}(a)). For the case of \sm{`p'}-polarized input laser at the resonance angle of $\theta_{inc}$=23$^{\circ}$, we can see a 3-fold increase in the electron flux for the grating target as compared to the flat target. More importantly, the electrons are preferentially accelerated along the positive surface direction (+90$^{\circ}$) for the grating target. We see a more isotropic electron angular distribution for the plain target. The \sm{`s'}-polarized pulse interaction with the gratings shows a 10-fold lower electron flux as compared with the \sm{`p'}-polarized pulse and it also shows no preferred direction for electron acceleration. The off-resonance case of $\theta_{inc}$=32$^{\circ}$ for the grating target (green curve in Fig. \ref{fig:exp_angdist_esm_combined} (a)) shows a slightly lower flux along both the surface directions of +90$^{\circ}$ and +270$^{\circ}$ indicating a lower surface coupling. The electron spectra for the grating target along three directions, {\it i.e.}, positive surface (+90$^{\circ}$), negative surface (+270$^{\circ}$) and specular (+23$^{\circ}$) reflection are shown in Fig. \ref{fig:exp_angdist_esm_combined}(c)-(d). Note that the cutoff energies and the temperature are higher along the positive surface direction, which agrees well with the angular distributions and indicates preferential electron acceleration along the positive surface direction. This however is in contrast to the surface plasma wave mode for a sub-$\lambda$ grating target, where the SPW fields and the electron acceleration are expected along the negative surface direction. For the plain target, the corresponding electron spectra are shown in \ref{fig:exp_angdist_esm_combined}(c) where we see similar electron cutoff energies and temperatures along all three directions, again indicating a more isotropic distribution of electrons. 
\smallskip

\renewcommand{\thefigure}{3}
\setlength{\belowcaptionskip}{-10pt}
\begin{figure}[!ht]
    \centering
    \includegraphics[width = \linewidth]{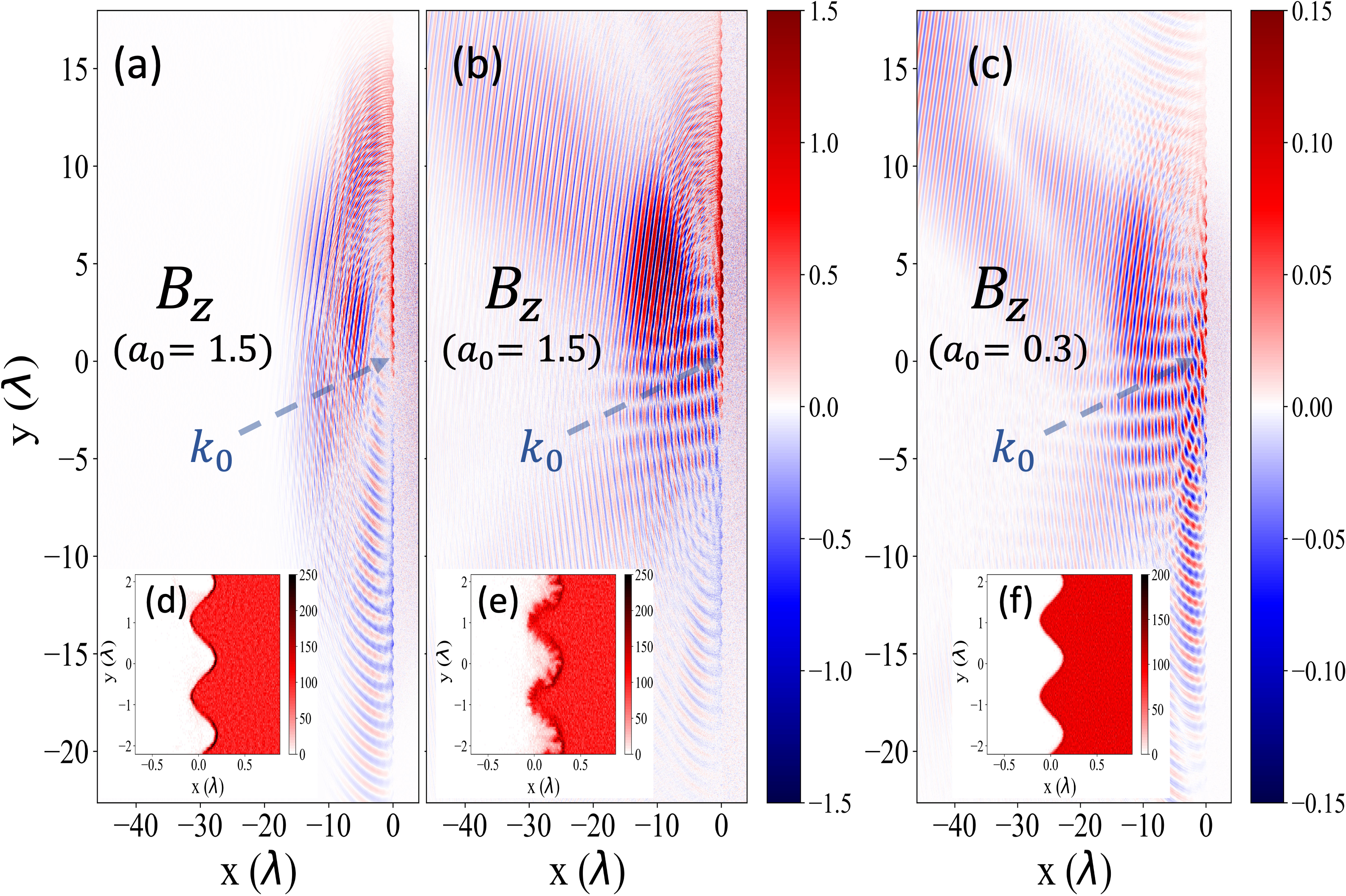}
    \caption{$B_z$ field profiles after reflection from the grating surface for (a) $a_0$=1.5, "clean" pulse, (b) $a_0$=1.5, "structured" pulse, (c) $a_0$=0.3, "structured" pulse. Figure insets (d)-(f) show the corresponding electron densities of the grating when the laser peak hits the target.}
    \label{fig:field_400nm}
\end{figure}

\renewcommand{\thefigure}{4}
\setlength{\belowcaptionskip}{-10pt}
\begin{figure*}[!ht]
    \centering
    \includegraphics[width = \linewidth]{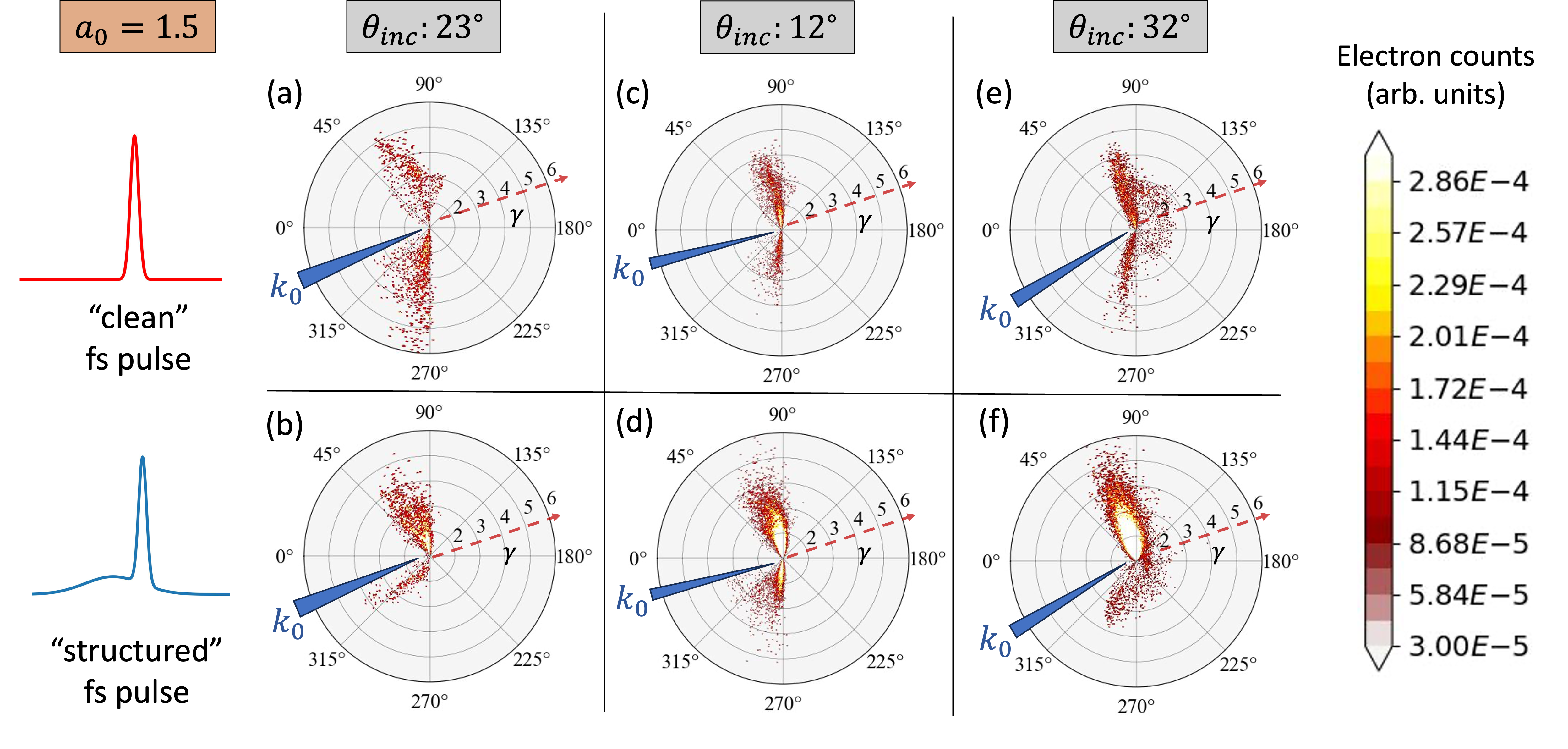}
    \caption{Electron distributions for the grating target for $a_0=1.5$ from the simulations. (a) $\theta_{inc}$=23$^{\circ}$, "clean" pulse, (b) $\theta_{inc}$=23$^{\circ}$, "structured" pulse, (c) $\theta_{inc}$=12$^{\circ}$, "clean" pulse, (d) $\theta_{inc}$=12$^{\circ}$, "structured" pulse,(e) $\theta_{inc}$=32$^{\circ}$, "clean" pulse, and (f) $\theta_{inc}$=32$^{\circ}$, "structured" pulse, respectively.   }
    \label{fig:ang_dist_sim}
\end{figure*}

 To understand the nature of the interaction clearly, we performed laser absorption measurements in two types of targets, a flat BK-7 glass target and a sub-$\lambda$ grating target, and the angle of incidence was chosen as 23$^{\circ}$ to resonantly excite surface plasma waves and enhance absorption in the latter. The reflected, transmitted and diffracted laser energies were fed to  OPHIR PE-50C and OPHIR PE-100C pyroelectric detectors to infer the laser absorption by the target.  A large input intensity range (three orders of magnitude, $10^{16}$ to $10^{19}$ W/cm$^2$) is covered to characterize the absorption. Fig. \ref{fig:exp_angdist_esm_combined}(b) shows that the absorption in a flat target increases from 30\% to 60\% with increasing input intensity. On the other hand, the grating target shows 80\% absorption over the entire intensity range of $10^{16}$-$10^{19}$ W/cm$^2$. The s-polarization absorption in the grating target is also higher than the flat target attributable to the geometric local electric-field enhancements in the vicinity of the structure \cite{Rajeev_OL2004}. 
\smallskip

To further understand the experimental results, 2D3V PIC simulations have been performed with the open-source code SMILEI \cite{Smilei_DEROUILLAT2018351}. The geometry of the plasma lies in the \sm{($x$, $y$)} plane for $x\ge 0$, its surface being along the \sm{$y$}-direction. The driving laser is a p-polarized Gaussian pulse with a waist equal to $5\lambda_0$ ($\sim2$ $\mu$m) and two temporal profiles were considered, "clean" and "structured" pulse (as depicted in Fig. \ref{fig:setup}(b)). The laser wavelength is chosen as 400 nm.  The laser pulse impinges on the plasma interface at an angle $\theta_{inc}$ to the normal surface along the \sm{$x$-}direction. The plasma grating has constant electron density $n_0$ with a sinusoidal-modulated vacuum–plasma interface located at $x(y) = (h/2) \sin (2\pi y/d)$ where $h$ is the grating depth and $d$ is the period. In all cases studied, we considered $d = 0.694$ $\lambda_0$ (= 277 nm, sub-$\lambda$ grating) and we used $h = 0.25$ $\lambda_0$ (= 0.1 $\mu$m) for the grating depth. For the flat target simulations, $d = 0$ was used, with all other parameters unchanged. The plasma consists of electrons with a small initial temperature of $T_e = 50$ eV as well as a neutralizing background of ions free to move in the space with initial temperature $T_i/(ZT_e) = 0.1$, where $Z=1$ is the atomic number. The electron density $n_0$ was set to 100 $n_c$ (critical density). The normalized vector potential, $a_0 = 1.5$ ($\sim 7.7 \times 10^{18}$ W/cm$^2$) was used in the simulations for electron angular distributions and spectra from the grating and flat targets. 
\smallskip

The laser field (B$_z$) profiles from the grating target are shown in Fig. \ref{fig:field_400nm}. These are shown at the time instant right after reflection of the input laser from the grating surface and the two cases correspond to the two temporal shapes considered in the simulations. For the $a_0=1.5$ case, we notice a sharp distinction between the two cases - (a) the "clean" pulse leads to efficient excitation of the surface plasma wave (SPW) mode, whereas (b) for the "structured" pulse, the SPW mode is suppressed.  On the other hand, for the (c) $a_0=0.3$ case, even with the "structured" pulse, we can efficiently drive the SPW modes. We also investigate the electron density profiles as given in Fig. \ref{fig:field_400nm} (d)-(f) right when the laser peak reaches them. Figure insets (d) and (e) depict the density profiles for the laser without and with the femtosecond pre-structures for $a_0=1.5$. We can see a significant alteration of the electron densities at the grating surface due to the femtosecond pre-structure, which in turn affects the field profiles, laser absorption and coupling mechanisms. Also, we note that the femtosecond-structure does not generate any additional pre-plasma scale length, nor change the height/period of the gratings, as these can also alter the resonance conditions and affect the coupling to the surface plasma waves. Rather, it produces a rippling effect on the electron densities, which affects the interactions. Figure \ref{fig:field_400nm} (f) shows the density profile for $a_0=0.3$ case with the structured pulse. The grating structure is preserved without any significant change in this case. This indicates that, at sub-relativistic intensities, the pulse structure does not play a significant role in the efficient generation of surface plasma waves. 
\smallskip

The angular distribution of electrons from simulations, as shown in Fig. \ref{fig:ang_dist_sim}, also shows a major effect of the femtosecond pre-structure on the directions, flux and energies of the electrons. These are plotted for three cases of the angle of incidence ($\theta_{inc}$=12$^{\circ}$/23$^{\circ}$/32$^{\circ}$), where $\theta_{inc}$=23$^{\circ}$ is the resonant case for surface plasma wave generation and the other two the non-resonant cases. Also, these are plotted considering electrons above 100 keV energy (as in the experiment). For the resonant case ({\it i.e.}, $\theta_{inc}$=23$^{\circ}$, the SPW mode along the negative surface (270$^{\circ}$) can be seen in case \ref{fig:ang_dist_sim}(a), which is for the "clean" pulse case, with their $\gamma$ reaching 6, which corresponds to $\sim$2.5 MeV, and the electrons along the positive surface (90$^{\circ}$ reach a maximum of $\sim$1.5 MeV. For the \ref{fig:ang_dist_sim}(b) case, which corresponds to the "structured" pulse, the electrons are preferentially accelerated along the positive surface with a maximum energy of $\sim$1.5 MeV and the SPW mode along the negative surface is suppressed with a maximum energy of $\sim$0.5 MeV. Also, we see a significantly higher flux of electrons towards 90$^{\circ}$ than at 270$^{\circ}$. For the non-resonant angles, as shown in \ref{fig:ang_dist_sim} (c)-(f), we see that the electron distributions and peak energies remain similar for both pulse shapes, but there is an increase in flux for the structured pulse case. This is because of the higher energy dump in the structured pulse case. Thus, the electron distributions and energy spectra for the grating resonance case are more sensitive to the pulse shape.
\smallskip

To conclude, we have performed experiments and simulations to explore the fundamental interactions of extreme-contrast 400 nm pulses experiments with flat and sub-$\lambda$ grating targets, and in particular explored the role of femtosecond structures in high-intensity, laser-solid interactions. We show that femtosecond pre-structures, which are inherently produced in efficient SHG of terawatt/petawatt class lasers modify \AC{the local electron densities, field profiles, and electron distributions with regards to} the SPW excitation in the grating target. We also presented that these may not play a significant role at sub-relativistic intensities but play a very significant role past the relativistic intensity barrier. We emphasize that the femtosecond temporal structures play a crucial role in such extreme contrast interactions with solids and must be carefully considered while understanding the interaction. We reaffirm that the modifications of the SPW response can serve as a sensitive indicator of the contrast at the femtosecond levels. \\

\textbf{Acknowledgement} \\
GRK acknowledges J.C. Bose Fellowship grant
(JBR/2020/000039) from the Science and Engineering
Board (SERB), Government of India. ADL acknowledges
partial support from the Infosys-TIFR Leading Edge Research Grant (Cycle 2). CA acknowledges partial support from the Sarojini Damodaran International Student Travel Fellowship. CA acknowledges Mr. Anilkumar Naik for his support in setting up and running SMILEI PIC simulations in the TIFR High Performance Computing (HPC) facility. Simulations were also performed on the Irene-SKL machine hosted at TGCC- France, using High Performance Computing resources from GENCI-TGCC (Grant No. 2021-x2016057678).

\bibliography{references}

\begin{thebibliography}{47}%
\makeatletter
\providecommand \@ifxundefined [1]{%
 \@ifx{#1\undefined}
}%
\providecommand \@ifnum [1]{%
 \ifnum #1\expandafter \@firstoftwo
 \else \expandafter \@secondoftwo
 \fi
}%
\providecommand \@ifx [1]{%
 \ifx #1\expandafter \@firstoftwo
 \else \expandafter \@secondoftwo
 \fi
}%
\providecommand \natexlab [1]{#1}%
\providecommand \enquote  [1]{``#1''}%
\providecommand \bibnamefont  [1]{#1}%
\providecommand \bibfnamefont [1]{#1}%
\providecommand \citenamefont [1]{#1}%
\providecommand \href@noop [0]{\@secondoftwo}%
\providecommand \href [0]{\begingroup \@sanitize@url \@href}%
\providecommand \@href[1]{\@@startlink{#1}\@@href}%
\providecommand \@@href[1]{\endgroup#1\@@endlink}%
\providecommand \@sanitize@url [0]{\catcode `\\12\catcode `\$12\catcode `\&12\catcode `\#12\catcode `\^12\catcode `\_12\catcode `\%12\relax}%
\providecommand \@@startlink[1]{}%
\providecommand \@@endlink[0]{}%
\providecommand \url  [0]{\begingroup\@sanitize@url \@url }%
\providecommand \@url [1]{\endgroup\@href {#1}{\urlprefix }}%
\providecommand \urlprefix  [0]{URL }%
\providecommand \Eprint [0]{\href }%
\providecommand \doibase [0]{https://doi.org/}%
\providecommand \selectlanguage [0]{\@gobble}%
\providecommand \bibinfo  [0]{\@secondoftwo}%
\providecommand \bibfield  [0]{\@secondoftwo}%
\providecommand \translation [1]{[#1]}%
\providecommand \BibitemOpen [0]{}%
\providecommand \bibitemStop [0]{}%
\providecommand \bibitemNoStop [0]{.\EOS\space}%
\providecommand \EOS [0]{\spacefactor3000\relax}%
\providecommand \BibitemShut  [1]{\csname bibitem#1\endcsname}%
\let\auto@bib@innerbib\@empty
\bibitem [{\citenamefont {Drake}(2006)}]{drake_2006}%
  \BibitemOpen
  \bibfield  {author} {\bibinfo {author} {\bibfnamefont {R.~P.}\ \bibnamefont {Drake}},\ }\href@noop {} {\emph {\bibinfo {title} {High Energy Density Physics}}}\ (\bibinfo  {publisher} {Springer-Verlag Berlin Heidelberg},\ \bibinfo {year} {2006})\BibitemShut {NoStop}%
\bibitem [{\citenamefont {Kaw}(2017)}]{Kaw2017}%
  \BibitemOpen
  \bibfield  {author} {\bibinfo {author} {\bibfnamefont {P.~K.}\ \bibnamefont {Kaw}},\ }\bibfield  {title} {\bibinfo {title} {Nonlinear laser--plasma interactions},\ }\href {https://doi.org/10.1007/s41614-017-0005-2} {\bibfield  {journal} {\bibinfo  {journal} {Reviews of Modern Plasma Physics}\ }\textbf {\bibinfo {volume} {1}},\ \bibinfo {pages} {2} (\bibinfo {year} {2017})}\BibitemShut {NoStop}%
\bibitem [{\citenamefont {Cerchez}\ \emph {et~al.}(2008)\citenamefont {Cerchez}, \citenamefont {Jung}, \citenamefont {Osterholz}, \citenamefont {Toncian}, \citenamefont {Willi}, \citenamefont {Mulser},\ and\ \citenamefont {Ruhl}}]{Cerchez_PhysRevLett.100.245001}%
  \BibitemOpen
  \bibfield  {author} {\bibinfo {author} {\bibfnamefont {M.}~\bibnamefont {Cerchez}}, \bibinfo {author} {\bibfnamefont {R.}~\bibnamefont {Jung}}, \bibinfo {author} {\bibfnamefont {J.}~\bibnamefont {Osterholz}}, \bibinfo {author} {\bibfnamefont {T.}~\bibnamefont {Toncian}}, \bibinfo {author} {\bibfnamefont {O.}~\bibnamefont {Willi}}, \bibinfo {author} {\bibfnamefont {P.}~\bibnamefont {Mulser}},\ and\ \bibinfo {author} {\bibfnamefont {H.}~\bibnamefont {Ruhl}},\ }\bibfield  {title} {\bibinfo {title} {Absorption of ultrashort laser pulses in strongly overdense targets},\ }\href {https://doi.org/10.1103/PhysRevLett.100.245001} {\bibfield  {journal} {\bibinfo  {journal} {Phys. Rev. Lett.}\ }\textbf {\bibinfo {volume} {100}},\ \bibinfo {pages} {245001} (\bibinfo {year} {2008})}\BibitemShut {NoStop}%
\bibitem [{\citenamefont {Ping}\ \emph {et~al.}(2008)\citenamefont {Ping}, \citenamefont {Shepherd}, \citenamefont {Lasinski}, \citenamefont {Tabak}, \citenamefont {Chen}, \citenamefont {Chung}, \citenamefont {Fournier}, \citenamefont {Hansen}, \citenamefont {Kemp}, \citenamefont {Liedahl}, \citenamefont {Widmann}, \citenamefont {Wilks}, \citenamefont {Rozmus},\ and\ \citenamefont {Sherlock}}]{Ping_PhysRevLett.100.085004}%
  \BibitemOpen
  \bibfield  {author} {\bibinfo {author} {\bibfnamefont {Y.}~\bibnamefont {Ping}}, \bibinfo {author} {\bibfnamefont {R.}~\bibnamefont {Shepherd}}, \bibinfo {author} {\bibfnamefont {B.~F.}\ \bibnamefont {Lasinski}}, \bibinfo {author} {\bibfnamefont {M.}~\bibnamefont {Tabak}}, \bibinfo {author} {\bibfnamefont {H.}~\bibnamefont {Chen}}, \bibinfo {author} {\bibfnamefont {H.~K.}\ \bibnamefont {Chung}}, \bibinfo {author} {\bibfnamefont {K.~B.}\ \bibnamefont {Fournier}}, \bibinfo {author} {\bibfnamefont {S.~B.}\ \bibnamefont {Hansen}}, \bibinfo {author} {\bibfnamefont {A.}~\bibnamefont {Kemp}}, \bibinfo {author} {\bibfnamefont {D.~A.}\ \bibnamefont {Liedahl}}, \bibinfo {author} {\bibfnamefont {K.}~\bibnamefont {Widmann}}, \bibinfo {author} {\bibfnamefont {S.~C.}\ \bibnamefont {Wilks}}, \bibinfo {author} {\bibfnamefont {W.}~\bibnamefont {Rozmus}},\ and\ \bibinfo {author} {\bibfnamefont {M.}~\bibnamefont {Sherlock}},\ }\bibfield  {title} {\bibinfo {title} {Absorption of short laser pulses on solid targets in the
  ultrarelativistic regime},\ }\href {https://doi.org/10.1103/PhysRevLett.100.085004} {\bibfield  {journal} {\bibinfo  {journal} {Phys. Rev. Lett.}\ }\textbf {\bibinfo {volume} {100}},\ \bibinfo {pages} {085004} (\bibinfo {year} {2008})}\BibitemShut {NoStop}%
\bibitem [{\citenamefont {Kemp}\ \emph {et~al.}(2008)\citenamefont {Kemp}, \citenamefont {Sentoku},\ and\ \citenamefont {Tabak}}]{Kemp_PhysRevLett.101.075004}%
  \BibitemOpen
  \bibfield  {author} {\bibinfo {author} {\bibfnamefont {A.~J.}\ \bibnamefont {Kemp}}, \bibinfo {author} {\bibfnamefont {Y.}~\bibnamefont {Sentoku}},\ and\ \bibinfo {author} {\bibfnamefont {M.}~\bibnamefont {Tabak}},\ }\bibfield  {title} {\bibinfo {title} {Hot-electron energy coupling in ultraintense laser-matter interaction},\ }\href {https://doi.org/10.1103/PhysRevLett.101.075004} {\bibfield  {journal} {\bibinfo  {journal} {Phys. Rev. Lett.}\ }\textbf {\bibinfo {volume} {101}},\ \bibinfo {pages} {075004} (\bibinfo {year} {2008})}\BibitemShut {NoStop}%
\bibitem [{\citenamefont {Peebles}\ \emph {et~al.}(2017)\citenamefont {Peebles}, \citenamefont {Wei}, \citenamefont {Arefiev}, \citenamefont {McGuffey}, \citenamefont {Stephens}, \citenamefont {Theobald}, \citenamefont {Haberberger}, \citenamefont {Jarrott}, \citenamefont {Link}, \citenamefont {Chen}, \citenamefont {McLean}, \citenamefont {Sorokovikova}, \citenamefont {Krasheninnikov},\ and\ \citenamefont {Beg}}]{Peebles_2017}%
  \BibitemOpen
  \bibfield  {author} {\bibinfo {author} {\bibfnamefont {J.}~\bibnamefont {Peebles}}, \bibinfo {author} {\bibfnamefont {M.~S.}\ \bibnamefont {Wei}}, \bibinfo {author} {\bibfnamefont {A.~V.}\ \bibnamefont {Arefiev}}, \bibinfo {author} {\bibfnamefont {C.}~\bibnamefont {McGuffey}}, \bibinfo {author} {\bibfnamefont {R.~B.}\ \bibnamefont {Stephens}}, \bibinfo {author} {\bibfnamefont {W.}~\bibnamefont {Theobald}}, \bibinfo {author} {\bibfnamefont {D.}~\bibnamefont {Haberberger}}, \bibinfo {author} {\bibfnamefont {L.~C.}\ \bibnamefont {Jarrott}}, \bibinfo {author} {\bibfnamefont {A.}~\bibnamefont {Link}}, \bibinfo {author} {\bibfnamefont {H.}~\bibnamefont {Chen}}, \bibinfo {author} {\bibfnamefont {H.~S.}\ \bibnamefont {McLean}}, \bibinfo {author} {\bibfnamefont {A.}~\bibnamefont {Sorokovikova}}, \bibinfo {author} {\bibfnamefont {S.}~\bibnamefont {Krasheninnikov}},\ and\ \bibinfo {author} {\bibfnamefont {F.~N.}\ \bibnamefont {Beg}},\ }\bibfield  {title} {\bibinfo {title} {Investigation of laser pulse length and
  pre-plasma scale length impact on hot electron generation on omega-ep},\ }\href {https://doi.org/10.1088/1367-2630/aa5a21} {\bibfield  {journal} {\bibinfo  {journal} {New Journal of Physics}\ }\textbf {\bibinfo {volume} {19}},\ \bibinfo {pages} {023008} (\bibinfo {year} {2017})}\BibitemShut {NoStop}%
\bibitem [{\citenamefont {Kahaly}\ \emph {et~al.}(2013)\citenamefont {Kahaly}, \citenamefont {Monchoc\'e}, \citenamefont {Vincenti}, \citenamefont {Dzelzainis}, \citenamefont {Dromey}, \citenamefont {Zepf}, \citenamefont {Martin},\ and\ \citenamefont {Qu\'er\'e}}]{Kahaly_PhysRevLett.110.175001}%
  \BibitemOpen
  \bibfield  {author} {\bibinfo {author} {\bibfnamefont {S.}~\bibnamefont {Kahaly}}, \bibinfo {author} {\bibfnamefont {S.}~\bibnamefont {Monchoc\'e}}, \bibinfo {author} {\bibfnamefont {H.}~\bibnamefont {Vincenti}}, \bibinfo {author} {\bibfnamefont {T.}~\bibnamefont {Dzelzainis}}, \bibinfo {author} {\bibfnamefont {B.}~\bibnamefont {Dromey}}, \bibinfo {author} {\bibfnamefont {M.}~\bibnamefont {Zepf}}, \bibinfo {author} {\bibfnamefont {P.}~\bibnamefont {Martin}},\ and\ \bibinfo {author} {\bibfnamefont {F.}~\bibnamefont {Qu\'er\'e}},\ }\bibfield  {title} {\bibinfo {title} {Direct observation of density-gradient effects in harmonic generation from plasma mirrors},\ }\href {https://doi.org/10.1103/PhysRevLett.110.175001} {\bibfield  {journal} {\bibinfo  {journal} {Phys. Rev. Lett.}\ }\textbf {\bibinfo {volume} {110}},\ \bibinfo {pages} {175001} (\bibinfo {year} {2013})}\BibitemShut {NoStop}%
\bibitem [{\citenamefont {Beier}\ \emph {et~al.}(2022)\citenamefont {Beier}, \citenamefont {Allison}, \citenamefont {Efthimion}, \citenamefont {Flippo}, \citenamefont {Gao}, \citenamefont {Hansen}, \citenamefont {Hill}, \citenamefont {Hollinger}, \citenamefont {Logantha}, \citenamefont {Musthafa}, \citenamefont {Nedbailo}, \citenamefont {Senthilkumaran}, \citenamefont {Shepherd}, \citenamefont {Shlyaptsev}, \citenamefont {Song}, \citenamefont {Wang}, \citenamefont {Dollar}, \citenamefont {Rocca},\ and\ \citenamefont {Hussein}}]{Beier2022}%
  \BibitemOpen
  \bibfield  {author} {\bibinfo {author} {\bibfnamefont {N.~F.}\ \bibnamefont {Beier}}, \bibinfo {author} {\bibfnamefont {H.}~\bibnamefont {Allison}}, \bibinfo {author} {\bibfnamefont {P.}~\bibnamefont {Efthimion}}, \bibinfo {author} {\bibfnamefont {K.~A.}\ \bibnamefont {Flippo}}, \bibinfo {author} {\bibfnamefont {L.}~\bibnamefont {Gao}}, \bibinfo {author} {\bibfnamefont {S.~B.}\ \bibnamefont {Hansen}}, \bibinfo {author} {\bibfnamefont {K.}~\bibnamefont {Hill}}, \bibinfo {author} {\bibfnamefont {R.}~\bibnamefont {Hollinger}}, \bibinfo {author} {\bibfnamefont {M.}~\bibnamefont {Logantha}}, \bibinfo {author} {\bibfnamefont {Y.}~\bibnamefont {Musthafa}}, \bibinfo {author} {\bibfnamefont {R.}~\bibnamefont {Nedbailo}}, \bibinfo {author} {\bibfnamefont {V.}~\bibnamefont {Senthilkumaran}}, \bibinfo {author} {\bibfnamefont {R.}~\bibnamefont {Shepherd}}, \bibinfo {author} {\bibfnamefont {V.~N.}\ \bibnamefont {Shlyaptsev}}, \bibinfo {author} {\bibfnamefont {H.}~\bibnamefont {Song}}, \bibinfo {author} {\bibfnamefont
  {S.}~\bibnamefont {Wang}}, \bibinfo {author} {\bibfnamefont {F.}~\bibnamefont {Dollar}}, \bibinfo {author} {\bibfnamefont {J.~J.}\ \bibnamefont {Rocca}},\ and\ \bibinfo {author} {\bibfnamefont {A.~E.}\ \bibnamefont {Hussein}},\ }\bibfield  {title} {\bibinfo {title} {Homogeneous, micron-scale high-energy-density matter generated by relativistic laser-solid interactions},\ }\href {https://doi.org/10.1103/PhysRevLett.129.135001} {\bibfield  {journal} {\bibinfo  {journal} {Phys. Rev. Lett.}\ }\textbf {\bibinfo {volume} {129}},\ \bibinfo {pages} {135001} (\bibinfo {year} {2022})}\BibitemShut {NoStop}%
\bibitem [{\citenamefont {Dromey}\ \emph {et~al.}(2006)\citenamefont {Dromey}, \citenamefont {Zepf}, \citenamefont {Gopal}, \citenamefont {Lancaster}, \citenamefont {Wei}, \citenamefont {Krushelnick}, \citenamefont {Tatarakis}, \citenamefont {Vakakis}, \citenamefont {Moustaizis}, \citenamefont {Kodama}, \citenamefont {Tampo}, \citenamefont {Stoeckl}, \citenamefont {Clarke}, \citenamefont {Habara}, \citenamefont {Neely}, \citenamefont {Karsch},\ and\ \citenamefont {Norreys}}]{Dromey2006}%
  \BibitemOpen
  \bibfield  {author} {\bibinfo {author} {\bibfnamefont {B.}~\bibnamefont {Dromey}}, \bibinfo {author} {\bibfnamefont {M.}~\bibnamefont {Zepf}}, \bibinfo {author} {\bibfnamefont {A.}~\bibnamefont {Gopal}}, \bibinfo {author} {\bibfnamefont {K.}~\bibnamefont {Lancaster}}, \bibinfo {author} {\bibfnamefont {M.~S.}\ \bibnamefont {Wei}}, \bibinfo {author} {\bibfnamefont {K.}~\bibnamefont {Krushelnick}}, \bibinfo {author} {\bibfnamefont {M.}~\bibnamefont {Tatarakis}}, \bibinfo {author} {\bibfnamefont {N.}~\bibnamefont {Vakakis}}, \bibinfo {author} {\bibfnamefont {S.}~\bibnamefont {Moustaizis}}, \bibinfo {author} {\bibfnamefont {R.}~\bibnamefont {Kodama}}, \bibinfo {author} {\bibfnamefont {M.}~\bibnamefont {Tampo}}, \bibinfo {author} {\bibfnamefont {C.}~\bibnamefont {Stoeckl}}, \bibinfo {author} {\bibfnamefont {R.}~\bibnamefont {Clarke}}, \bibinfo {author} {\bibfnamefont {H.}~\bibnamefont {Habara}}, \bibinfo {author} {\bibfnamefont {D.}~\bibnamefont {Neely}}, \bibinfo {author} {\bibfnamefont {S.}~\bibnamefont
  {Karsch}},\ and\ \bibinfo {author} {\bibfnamefont {P.}~\bibnamefont {Norreys}},\ }\bibfield  {title} {\bibinfo {title} {High harmonic generation in the relativistic limit},\ }\href {https://doi.org/10.1038/nphys338} {\bibfield  {journal} {\bibinfo  {journal} {Nature Physics}\ }\textbf {\bibinfo {volume} {2}},\ \bibinfo {pages} {456} (\bibinfo {year} {2006})}\BibitemShut {NoStop}%
\bibitem [{\citenamefont {Dollar}\ \emph {et~al.}(2013)\citenamefont {Dollar}, \citenamefont {Cummings}, \citenamefont {Chvykov}, \citenamefont {Willingale}, \citenamefont {Vargas}, \citenamefont {Yanovsky}, \citenamefont {Zulick}, \citenamefont {Maksimchuk}, \citenamefont {Thomas},\ and\ \citenamefont {Krushelnick}}]{Dollar2013_HHG_Solids_PhysRevLett.110.175002}%
  \BibitemOpen
  \bibfield  {author} {\bibinfo {author} {\bibfnamefont {F.}~\bibnamefont {Dollar}}, \bibinfo {author} {\bibfnamefont {P.}~\bibnamefont {Cummings}}, \bibinfo {author} {\bibfnamefont {V.}~\bibnamefont {Chvykov}}, \bibinfo {author} {\bibfnamefont {L.}~\bibnamefont {Willingale}}, \bibinfo {author} {\bibfnamefont {M.}~\bibnamefont {Vargas}}, \bibinfo {author} {\bibfnamefont {V.}~\bibnamefont {Yanovsky}}, \bibinfo {author} {\bibfnamefont {C.}~\bibnamefont {Zulick}}, \bibinfo {author} {\bibfnamefont {A.}~\bibnamefont {Maksimchuk}}, \bibinfo {author} {\bibfnamefont {A.~G.~R.}\ \bibnamefont {Thomas}},\ and\ \bibinfo {author} {\bibfnamefont {K.}~\bibnamefont {Krushelnick}},\ }\bibfield  {title} {\bibinfo {title} {Scaling high-order harmonic generation from laser-solid interactions to ultrahigh intensity},\ }\href {https://doi.org/10.1103/PhysRevLett.110.175002} {\bibfield  {journal} {\bibinfo  {journal} {Phys. Rev. Lett.}\ }\textbf {\bibinfo {volume} {110}},\ \bibinfo {pages} {175002} (\bibinfo {year}
  {2013})}\BibitemShut {NoStop}%
\bibitem [{\citenamefont {Singh}\ \emph {et~al.}(2022)\citenamefont {Singh}, \citenamefont {Li}, \citenamefont {Huang}, \citenamefont {Moreau}, \citenamefont {Hollinger}, \citenamefont {Junghans}, \citenamefont {Favalli}, \citenamefont {Calvi}, \citenamefont {Wang}, \citenamefont {Wang}, \citenamefont {Song}, \citenamefont {Rocca}, \citenamefont {Reinovsky},\ and\ \citenamefont {Palaniyappan}}]{Singh2022}%
  \BibitemOpen
  \bibfield  {author} {\bibinfo {author} {\bibfnamefont {P.~K.}\ \bibnamefont {Singh}}, \bibinfo {author} {\bibfnamefont {F.-Y.}\ \bibnamefont {Li}}, \bibinfo {author} {\bibfnamefont {C.-K.}\ \bibnamefont {Huang}}, \bibinfo {author} {\bibfnamefont {A.}~\bibnamefont {Moreau}}, \bibinfo {author} {\bibfnamefont {R.}~\bibnamefont {Hollinger}}, \bibinfo {author} {\bibfnamefont {A.}~\bibnamefont {Junghans}}, \bibinfo {author} {\bibfnamefont {A.}~\bibnamefont {Favalli}}, \bibinfo {author} {\bibfnamefont {C.}~\bibnamefont {Calvi}}, \bibinfo {author} {\bibfnamefont {S.}~\bibnamefont {Wang}}, \bibinfo {author} {\bibfnamefont {Y.}~\bibnamefont {Wang}}, \bibinfo {author} {\bibfnamefont {H.}~\bibnamefont {Song}}, \bibinfo {author} {\bibfnamefont {J.~J.}\ \bibnamefont {Rocca}}, \bibinfo {author} {\bibfnamefont {R.~E.}\ \bibnamefont {Reinovsky}},\ and\ \bibinfo {author} {\bibfnamefont {S.}~\bibnamefont {Palaniyappan}},\ }\bibfield  {title} {\bibinfo {title} {Vacuum laser acceleration of super-ponderomotive electrons using
  relativistic transparency injection},\ }\href {https://doi.org/10.1038/s41467-021-27691-w} {\bibfield  {journal} {\bibinfo  {journal} {Nature Communications}\ }\textbf {\bibinfo {volume} {13}},\ \bibinfo {pages} {54} (\bibinfo {year} {2022})}\BibitemShut {NoStop}%
\bibitem [{\citenamefont {Macchi}\ \emph {et~al.}(2010)\citenamefont {Macchi}, \citenamefont {Veghini}, \citenamefont {Liseykina},\ and\ \citenamefont {Pegoraro}}]{Macchi_2010_RPA}%
  \BibitemOpen
  \bibfield  {author} {\bibinfo {author} {\bibfnamefont {A.}~\bibnamefont {Macchi}}, \bibinfo {author} {\bibfnamefont {S.}~\bibnamefont {Veghini}}, \bibinfo {author} {\bibfnamefont {T.~V.}\ \bibnamefont {Liseykina}},\ and\ \bibinfo {author} {\bibfnamefont {F.}~\bibnamefont {Pegoraro}},\ }\bibfield  {title} {\bibinfo {title} {Radiation pressure acceleration of ultrathin foils},\ }\href {https://doi.org/10.1088/1367-2630/12/4/045013} {\bibfield  {journal} {\bibinfo  {journal} {New Journal of Physics}\ }\textbf {\bibinfo {volume} {12}},\ \bibinfo {pages} {045013} (\bibinfo {year} {2010})}\BibitemShut {NoStop}%
\bibitem [{\citenamefont {Henig}\ \emph {et~al.}(2009)\citenamefont {Henig}, \citenamefont {Steinke}, \citenamefont {Schn\"urer}, \citenamefont {Sokollik}, \citenamefont {H\"orlein}, \citenamefont {Kiefer}, \citenamefont {Jung}, \citenamefont {Schreiber}, \citenamefont {Hegelich}, \citenamefont {Yan}, \citenamefont {Meyer-ter Vehn}, \citenamefont {Tajima}, \citenamefont {Nickles}, \citenamefont {Sandner},\ and\ \citenamefont {Habs}}]{Henig_2009_PhysRevLett.103.245003_RPA}%
  \BibitemOpen
  \bibfield  {author} {\bibinfo {author} {\bibfnamefont {A.}~\bibnamefont {Henig}}, \bibinfo {author} {\bibfnamefont {S.}~\bibnamefont {Steinke}}, \bibinfo {author} {\bibfnamefont {M.}~\bibnamefont {Schn\"urer}}, \bibinfo {author} {\bibfnamefont {T.}~\bibnamefont {Sokollik}}, \bibinfo {author} {\bibfnamefont {R.}~\bibnamefont {H\"orlein}}, \bibinfo {author} {\bibfnamefont {D.}~\bibnamefont {Kiefer}}, \bibinfo {author} {\bibfnamefont {D.}~\bibnamefont {Jung}}, \bibinfo {author} {\bibfnamefont {J.}~\bibnamefont {Schreiber}}, \bibinfo {author} {\bibfnamefont {B.~M.}\ \bibnamefont {Hegelich}}, \bibinfo {author} {\bibfnamefont {X.~Q.}\ \bibnamefont {Yan}}, \bibinfo {author} {\bibfnamefont {J.}~\bibnamefont {Meyer-ter Vehn}}, \bibinfo {author} {\bibfnamefont {T.}~\bibnamefont {Tajima}}, \bibinfo {author} {\bibfnamefont {P.~V.}\ \bibnamefont {Nickles}}, \bibinfo {author} {\bibfnamefont {W.}~\bibnamefont {Sandner}},\ and\ \bibinfo {author} {\bibfnamefont {D.}~\bibnamefont {Habs}},\ }\bibfield  {title} {\bibinfo
  {title} {Radiation-pressure acceleration of ion beams driven by circularly polarized laser pulses},\ }\href {https://doi.org/10.1103/PhysRevLett.103.245003} {\bibfield  {journal} {\bibinfo  {journal} {Phys. Rev. Lett.}\ }\textbf {\bibinfo {volume} {103}},\ \bibinfo {pages} {245003} (\bibinfo {year} {2009})}\BibitemShut {NoStop}%
\bibitem [{\citenamefont {Purvis}\ \emph {et~al.}(2013)\citenamefont {Purvis}, \citenamefont {Shlyaptsev}, \citenamefont {Hollinger}, \citenamefont {Bargsten}, \citenamefont {Pukhov}, \citenamefont {Prieto}, \citenamefont {Wang}, \citenamefont {Luther}, \citenamefont {Yin}, \citenamefont {Wang},\ and\ \citenamefont {Rocca}}]{Purvis2013}%
  \BibitemOpen
  \bibfield  {author} {\bibinfo {author} {\bibfnamefont {M.~A.}\ \bibnamefont {Purvis}}, \bibinfo {author} {\bibfnamefont {V.~N.}\ \bibnamefont {Shlyaptsev}}, \bibinfo {author} {\bibfnamefont {R.}~\bibnamefont {Hollinger}}, \bibinfo {author} {\bibfnamefont {C.}~\bibnamefont {Bargsten}}, \bibinfo {author} {\bibfnamefont {A.}~\bibnamefont {Pukhov}}, \bibinfo {author} {\bibfnamefont {A.}~\bibnamefont {Prieto}}, \bibinfo {author} {\bibfnamefont {Y.}~\bibnamefont {Wang}}, \bibinfo {author} {\bibfnamefont {B.~M.}\ \bibnamefont {Luther}}, \bibinfo {author} {\bibfnamefont {L.}~\bibnamefont {Yin}}, \bibinfo {author} {\bibfnamefont {S.}~\bibnamefont {Wang}},\ and\ \bibinfo {author} {\bibfnamefont {J.~J.}\ \bibnamefont {Rocca}},\ }\bibfield  {title} {\bibinfo {title} {Relativistic plasma nanophotonics for ultrahigh energy density physics},\ }\href {https://doi.org/10.1038/nphoton.2013.217} {\bibfield  {journal} {\bibinfo  {journal} {Nature Photonics}\ }\textbf {\bibinfo {volume} {7}},\ \bibinfo {pages} {796} (\bibinfo
  {year} {2013})}\BibitemShut {NoStop}%
\bibitem [{\citenamefont {Cantono}\ \emph {et~al.}(2018)\citenamefont {Cantono}, \citenamefont {Fedeli}, \citenamefont {Sgattoni}, \citenamefont {Denoeud}, \citenamefont {Chopineau}, \citenamefont {R\'eau}, \citenamefont {Ceccotti},\ and\ \citenamefont {Macchi}}]{Cantono_PRL2018}%
  \BibitemOpen
  \bibfield  {author} {\bibinfo {author} {\bibfnamefont {G.}~\bibnamefont {Cantono}}, \bibinfo {author} {\bibfnamefont {L.}~\bibnamefont {Fedeli}}, \bibinfo {author} {\bibfnamefont {A.}~\bibnamefont {Sgattoni}}, \bibinfo {author} {\bibfnamefont {A.}~\bibnamefont {Denoeud}}, \bibinfo {author} {\bibfnamefont {L.}~\bibnamefont {Chopineau}}, \bibinfo {author} {\bibfnamefont {F.}~\bibnamefont {R\'eau}}, \bibinfo {author} {\bibfnamefont {T.}~\bibnamefont {Ceccotti}},\ and\ \bibinfo {author} {\bibfnamefont {A.}~\bibnamefont {Macchi}},\ }\bibfield  {title} {\bibinfo {title} {Extreme ultraviolet beam enhancement by relativistic surface plasmons},\ }\href {https://doi.org/10.1103/PhysRevLett.120.264803} {\bibfield  {journal} {\bibinfo  {journal} {Phys. Rev. Lett.}\ }\textbf {\bibinfo {volume} {120}},\ \bibinfo {pages} {264803} (\bibinfo {year} {2018})}\BibitemShut {NoStop}%
\bibitem [{\citenamefont {Fedeli}\ \emph {et~al.}(2016)\citenamefont {Fedeli}, \citenamefont {Sgattoni}, \citenamefont {Cantono}, \citenamefont {Garzella}, \citenamefont {R\'eau}, \citenamefont {Prencipe}, \citenamefont {Passoni}, \citenamefont {Raynaud}, \citenamefont {Kv\ifmmode \check{e}\else \v{e}\fi{}to\ifmmode~\check{n}\else \v{n}\fi{}}, \citenamefont {Proska}, \citenamefont {Macchi},\ and\ \citenamefont {Ceccotti}}]{Fedeli_PRL_2016}%
  \BibitemOpen
  \bibfield  {author} {\bibinfo {author} {\bibfnamefont {L.}~\bibnamefont {Fedeli}}, \bibinfo {author} {\bibfnamefont {A.}~\bibnamefont {Sgattoni}}, \bibinfo {author} {\bibfnamefont {G.}~\bibnamefont {Cantono}}, \bibinfo {author} {\bibfnamefont {D.}~\bibnamefont {Garzella}}, \bibinfo {author} {\bibfnamefont {F.}~\bibnamefont {R\'eau}}, \bibinfo {author} {\bibfnamefont {I.}~\bibnamefont {Prencipe}}, \bibinfo {author} {\bibfnamefont {M.}~\bibnamefont {Passoni}}, \bibinfo {author} {\bibfnamefont {M.}~\bibnamefont {Raynaud}}, \bibinfo {author} {\bibfnamefont {M.}~\bibnamefont {Kv\ifmmode \check{e}\else \v{e}\fi{}to\ifmmode~\check{n}\else \v{n}\fi{}}}, \bibinfo {author} {\bibfnamefont {J.}~\bibnamefont {Proska}}, \bibinfo {author} {\bibfnamefont {A.}~\bibnamefont {Macchi}},\ and\ \bibinfo {author} {\bibfnamefont {T.}~\bibnamefont {Ceccotti}},\ }\bibfield  {title} {\bibinfo {title} {Electron acceleration by relativistic surface plasmons in laser-grating interaction},\ }\href
  {https://doi.org/10.1103/PhysRevLett.116.015001} {\bibfield  {journal} {\bibinfo  {journal} {Phys. Rev. Lett.}\ }\textbf {\bibinfo {volume} {116}},\ \bibinfo {pages} {015001} (\bibinfo {year} {2016})}\BibitemShut {NoStop}%
\bibitem [{\citenamefont {Ceccotti}\ \emph {et~al.}(2013)\citenamefont {Ceccotti}, \citenamefont {Floquet}, \citenamefont {Sgattoni}, \citenamefont {Bigongiari}, \citenamefont {Klimo}, \citenamefont {Raynaud}, \citenamefont {Riconda}, \citenamefont {Heron}, \citenamefont {Baffigi}, \citenamefont {Labate}, \citenamefont {Gizzi}, \citenamefont {Vassura}, \citenamefont {Fuchs}, \citenamefont {Passoni}, \citenamefont {Kv\ifmmode~\check{e}\else \v{e}\fi{}ton}, \citenamefont {Novotny}, \citenamefont {Possolt}, \citenamefont {Prok\ifmmode~\mathring{u}\else \r{u}\fi{}pek}, \citenamefont {Pro\ifmmode~\check{s}\else \v{s}\fi{}ka}, \citenamefont {P\ifmmode~\check{s}\else \v{s}\fi{}ikal}, \citenamefont {\ifmmode~\check{S}\else \v{S}\fi{}tolcov\'a}, \citenamefont {Velyhan}, \citenamefont {Bougeard}, \citenamefont {D'Oliveira}, \citenamefont {Tcherbakoff}, \citenamefont {R\'eau}, \citenamefont {Martin},\ and\ \citenamefont {Macchi}}]{Ceccotti_PRL2013}%
  \BibitemOpen
  \bibfield  {author} {\bibinfo {author} {\bibfnamefont {T.}~\bibnamefont {Ceccotti}}, \bibinfo {author} {\bibfnamefont {V.}~\bibnamefont {Floquet}}, \bibinfo {author} {\bibfnamefont {A.}~\bibnamefont {Sgattoni}}, \bibinfo {author} {\bibfnamefont {A.}~\bibnamefont {Bigongiari}}, \bibinfo {author} {\bibfnamefont {O.}~\bibnamefont {Klimo}}, \bibinfo {author} {\bibfnamefont {M.}~\bibnamefont {Raynaud}}, \bibinfo {author} {\bibfnamefont {C.}~\bibnamefont {Riconda}}, \bibinfo {author} {\bibfnamefont {A.}~\bibnamefont {Heron}}, \bibinfo {author} {\bibfnamefont {F.}~\bibnamefont {Baffigi}}, \bibinfo {author} {\bibfnamefont {L.}~\bibnamefont {Labate}}, \bibinfo {author} {\bibfnamefont {L.~A.}\ \bibnamefont {Gizzi}}, \bibinfo {author} {\bibfnamefont {L.}~\bibnamefont {Vassura}}, \bibinfo {author} {\bibfnamefont {J.}~\bibnamefont {Fuchs}}, \bibinfo {author} {\bibfnamefont {M.}~\bibnamefont {Passoni}}, \bibinfo {author} {\bibfnamefont {M.}~\bibnamefont {Kv\ifmmode~\check{e}\else \v{e}\fi{}ton}}, \bibinfo {author}
  {\bibfnamefont {F.}~\bibnamefont {Novotny}}, \bibinfo {author} {\bibfnamefont {M.}~\bibnamefont {Possolt}}, \bibinfo {author} {\bibfnamefont {J.}~\bibnamefont {Prok\ifmmode~\mathring{u}\else \r{u}\fi{}pek}}, \bibinfo {author} {\bibfnamefont {J.}~\bibnamefont {Pro\ifmmode~\check{s}\else \v{s}\fi{}ka}}, \bibinfo {author} {\bibfnamefont {J.}~\bibnamefont {P\ifmmode~\check{s}\else \v{s}\fi{}ikal}}, \bibinfo {author} {\bibfnamefont {L.}~\bibnamefont {\ifmmode~\check{S}\else \v{S}\fi{}tolcov\'a}}, \bibinfo {author} {\bibfnamefont {A.}~\bibnamefont {Velyhan}}, \bibinfo {author} {\bibfnamefont {M.}~\bibnamefont {Bougeard}}, \bibinfo {author} {\bibfnamefont {P.}~\bibnamefont {D'Oliveira}}, \bibinfo {author} {\bibfnamefont {O.}~\bibnamefont {Tcherbakoff}}, \bibinfo {author} {\bibfnamefont {F.}~\bibnamefont {R\'eau}}, \bibinfo {author} {\bibfnamefont {P.}~\bibnamefont {Martin}},\ and\ \bibinfo {author} {\bibfnamefont {A.}~\bibnamefont {Macchi}},\ }\bibfield  {title} {\bibinfo {title} {Evidence of resonant surface-wave
  excitation in the relativistic regime through measurements of proton acceleration from grating targets},\ }\href {https://doi.org/10.1103/PhysRevLett.111.185001} {\bibfield  {journal} {\bibinfo  {journal} {Phys. Rev. Lett.}\ }\textbf {\bibinfo {volume} {111}},\ \bibinfo {pages} {185001} (\bibinfo {year} {2013})}\BibitemShut {NoStop}%
\bibitem [{\citenamefont {Rajeev}\ \emph {et~al.}(2003)\citenamefont {Rajeev}, \citenamefont {Taneja}, \citenamefont {Ayyub}, \citenamefont {Sandhu},\ and\ \citenamefont {Kumar}}]{Rajeev-PRL2003}%
  \BibitemOpen
  \bibfield  {author} {\bibinfo {author} {\bibfnamefont {P.~P.}\ \bibnamefont {Rajeev}}, \bibinfo {author} {\bibfnamefont {P.}~\bibnamefont {Taneja}}, \bibinfo {author} {\bibfnamefont {P.}~\bibnamefont {Ayyub}}, \bibinfo {author} {\bibfnamefont {A.~S.}\ \bibnamefont {Sandhu}},\ and\ \bibinfo {author} {\bibfnamefont {G.~R.}\ \bibnamefont {Kumar}},\ }\bibfield  {title} {\bibinfo {title} {Metal nanoplasmas as bright sources of hard x-ray pulses},\ }\href {https://doi.org/10.1103/PhysRevLett.90.115002} {\bibfield  {journal} {\bibinfo  {journal} {Phys. Rev. Lett.}\ }\textbf {\bibinfo {volume} {90}},\ \bibinfo {pages} {115002} (\bibinfo {year} {2003})}\BibitemShut {NoStop}%
\bibitem [{\citenamefont {Bargsten}\ \emph {et~al.}(2017)\citenamefont {Bargsten}, \citenamefont {Hollinger}, \citenamefont {Capeluto}, \citenamefont {Kaymak}, \citenamefont {Pukhov}, \citenamefont {Wang}, \citenamefont {Rockwood}, \citenamefont {Wang}, \citenamefont {Keiss}, \citenamefont {Tommasini}, \citenamefont {London}, \citenamefont {Park}, \citenamefont {Busquet}, \citenamefont {Klapisch}, \citenamefont {Shlyaptsev},\ and\ \citenamefont {Rocca}}]{Rocca_SciAdv}%
  \BibitemOpen
  \bibfield  {author} {\bibinfo {author} {\bibfnamefont {C.}~\bibnamefont {Bargsten}}, \bibinfo {author} {\bibfnamefont {R.}~\bibnamefont {Hollinger}}, \bibinfo {author} {\bibfnamefont {M.~G.}\ \bibnamefont {Capeluto}}, \bibinfo {author} {\bibfnamefont {V.}~\bibnamefont {Kaymak}}, \bibinfo {author} {\bibfnamefont {A.}~\bibnamefont {Pukhov}}, \bibinfo {author} {\bibfnamefont {S.}~\bibnamefont {Wang}}, \bibinfo {author} {\bibfnamefont {A.}~\bibnamefont {Rockwood}}, \bibinfo {author} {\bibfnamefont {Y.}~\bibnamefont {Wang}}, \bibinfo {author} {\bibfnamefont {D.}~\bibnamefont {Keiss}}, \bibinfo {author} {\bibfnamefont {R.}~\bibnamefont {Tommasini}}, \bibinfo {author} {\bibfnamefont {R.}~\bibnamefont {London}}, \bibinfo {author} {\bibfnamefont {J.}~\bibnamefont {Park}}, \bibinfo {author} {\bibfnamefont {M.}~\bibnamefont {Busquet}}, \bibinfo {author} {\bibfnamefont {M.}~\bibnamefont {Klapisch}}, \bibinfo {author} {\bibfnamefont {V.~N.}\ \bibnamefont {Shlyaptsev}},\ and\ \bibinfo {author} {\bibfnamefont {J.~J.}\
  \bibnamefont {Rocca}},\ }\bibfield  {title} {\bibinfo {title} {Energy penetration into arrays of aligned nanowires irradiated with relativistic intensities: Scaling to terabar pressures},\ }\bibfield  {journal} {\bibinfo  {journal} {Science Advances}\ }\textbf {\bibinfo {volume} {3}},\ \href {https://doi.org/10.1126/sciadv.1601558} {10.1126/sciadv.1601558} (\bibinfo {year} {2017})\BibitemShut {NoStop}%
\bibitem [{\citenamefont {Kahaly}\ \emph {et~al.}(2008)\citenamefont {Kahaly}, \citenamefont {Yadav}, \citenamefont {Wang}, \citenamefont {Sengupta}, \citenamefont {Sheng}, \citenamefont {Das}, \citenamefont {Kaw},\ and\ \citenamefont {Kumar}}]{Kahaly-PRL2008}%
  \BibitemOpen
  \bibfield  {author} {\bibinfo {author} {\bibfnamefont {S.}~\bibnamefont {Kahaly}}, \bibinfo {author} {\bibfnamefont {S.~K.}\ \bibnamefont {Yadav}}, \bibinfo {author} {\bibfnamefont {W.~M.}\ \bibnamefont {Wang}}, \bibinfo {author} {\bibfnamefont {S.}~\bibnamefont {Sengupta}}, \bibinfo {author} {\bibfnamefont {Z.~M.}\ \bibnamefont {Sheng}}, \bibinfo {author} {\bibfnamefont {A.}~\bibnamefont {Das}}, \bibinfo {author} {\bibfnamefont {P.~K.}\ \bibnamefont {Kaw}},\ and\ \bibinfo {author} {\bibfnamefont {G.~R.}\ \bibnamefont {Kumar}},\ }\bibfield  {title} {\bibinfo {title} {Near-complete absorption of intense, ultrashort laser light by sub-$\ensuremath{\lambda}$ gratings},\ }\href {https://doi.org/10.1103/PhysRevLett.101.145001} {\bibfield  {journal} {\bibinfo  {journal} {Phys. Rev. Lett.}\ }\textbf {\bibinfo {volume} {101}},\ \bibinfo {pages} {145001} (\bibinfo {year} {2008})}\BibitemShut {NoStop}%
\bibitem [{\citenamefont {Lad}\ \emph {et~al.}(2022)\citenamefont {Lad}, \citenamefont {Mishima}, \citenamefont {Singh}, \citenamefont {Li}, \citenamefont {Adak}, \citenamefont {Chatterjee}, \citenamefont {Brijesh}, \citenamefont {Dalui}, \citenamefont {Inoue}, \citenamefont {Jha}, \citenamefont {Tata}, \citenamefont {Trivikram}, \citenamefont {Krishnamurthy}, \citenamefont {Chen}, \citenamefont {Sheng}, \citenamefont {Tanaka}, \citenamefont {Kumar},\ and\ \citenamefont {Habara}}]{Lad2022}%
  \BibitemOpen
  \bibfield  {author} {\bibinfo {author} {\bibfnamefont {A.~D.}\ \bibnamefont {Lad}}, \bibinfo {author} {\bibfnamefont {Y.}~\bibnamefont {Mishima}}, \bibinfo {author} {\bibfnamefont {P.~K.}\ \bibnamefont {Singh}}, \bibinfo {author} {\bibfnamefont {B.}~\bibnamefont {Li}}, \bibinfo {author} {\bibfnamefont {A.}~\bibnamefont {Adak}}, \bibinfo {author} {\bibfnamefont {G.}~\bibnamefont {Chatterjee}}, \bibinfo {author} {\bibfnamefont {P.}~\bibnamefont {Brijesh}}, \bibinfo {author} {\bibfnamefont {M.}~\bibnamefont {Dalui}}, \bibinfo {author} {\bibfnamefont {M.}~\bibnamefont {Inoue}}, \bibinfo {author} {\bibfnamefont {J.}~\bibnamefont {Jha}}, \bibinfo {author} {\bibfnamefont {S.}~\bibnamefont {Tata}}, \bibinfo {author} {\bibfnamefont {M.}~\bibnamefont {Trivikram}}, \bibinfo {author} {\bibfnamefont {M.}~\bibnamefont {Krishnamurthy}}, \bibinfo {author} {\bibfnamefont {M.}~\bibnamefont {Chen}}, \bibinfo {author} {\bibfnamefont {Z.~M.}\ \bibnamefont {Sheng}}, \bibinfo {author} {\bibfnamefont {K.~A.}\ \bibnamefont
  {Tanaka}}, \bibinfo {author} {\bibfnamefont {G.~R.}\ \bibnamefont {Kumar}},\ and\ \bibinfo {author} {\bibfnamefont {H.}~\bibnamefont {Habara}},\ }\bibfield  {title} {\bibinfo {title} {Luminous, relativistic, directional electron bunches from an intense laser driven grating plasma},\ }\href {https://doi.org/10.1038/s41598-022-21210-7} {\bibfield  {journal} {\bibinfo  {journal} {Scientific Reports}\ }\textbf {\bibinfo {volume} {12}},\ \bibinfo {pages} {16818} (\bibinfo {year} {2022})}\BibitemShut {NoStop}%
\bibitem [{\citenamefont {Price}\ \emph {et~al.}(1995)\citenamefont {Price}, \citenamefont {More}, \citenamefont {Walling}, \citenamefont {Guethlein}, \citenamefont {Shepherd}, \citenamefont {Stewart},\ and\ \citenamefont {White}}]{Price_PRL1995}%
  \BibitemOpen
  \bibfield  {author} {\bibinfo {author} {\bibfnamefont {D.~F.}\ \bibnamefont {Price}}, \bibinfo {author} {\bibfnamefont {R.~M.}\ \bibnamefont {More}}, \bibinfo {author} {\bibfnamefont {R.~S.}\ \bibnamefont {Walling}}, \bibinfo {author} {\bibfnamefont {G.}~\bibnamefont {Guethlein}}, \bibinfo {author} {\bibfnamefont {R.~L.}\ \bibnamefont {Shepherd}}, \bibinfo {author} {\bibfnamefont {R.~E.}\ \bibnamefont {Stewart}},\ and\ \bibinfo {author} {\bibfnamefont {W.~E.}\ \bibnamefont {White}},\ }\bibfield  {title} {\bibinfo {title} {Absorption of ultrashort laser pulses by solid targets heated rapidly to temperatures 1--1000 ev},\ }\href {https://doi.org/10.1103/PhysRevLett.75.252} {\bibfield  {journal} {\bibinfo  {journal} {Phys. Rev. Lett.}\ }\textbf {\bibinfo {volume} {75}},\ \bibinfo {pages} {252} (\bibinfo {year} {1995})}\BibitemShut {NoStop}%
\bibitem [{\citenamefont {Chen}\ \emph {et~al.}(2008)\citenamefont {Chen}, \citenamefont {Kando}, \citenamefont {Xu}, \citenamefont {Li}, \citenamefont {Koga}, \citenamefont {Chen}, \citenamefont {Xu}, \citenamefont {Yuan}, \citenamefont {Dong}, \citenamefont {Sheng}, \citenamefont {Bulanov}, \citenamefont {Kato}, \citenamefont {Zhang},\ and\ \citenamefont {Tajima}}]{Chen_PhysRevLett.100.045004}%
  \BibitemOpen
  \bibfield  {author} {\bibinfo {author} {\bibfnamefont {L.~M.}\ \bibnamefont {Chen}}, \bibinfo {author} {\bibfnamefont {M.}~\bibnamefont {Kando}}, \bibinfo {author} {\bibfnamefont {M.~H.}\ \bibnamefont {Xu}}, \bibinfo {author} {\bibfnamefont {Y.~T.}\ \bibnamefont {Li}}, \bibinfo {author} {\bibfnamefont {J.}~\bibnamefont {Koga}}, \bibinfo {author} {\bibfnamefont {M.}~\bibnamefont {Chen}}, \bibinfo {author} {\bibfnamefont {H.}~\bibnamefont {Xu}}, \bibinfo {author} {\bibfnamefont {X.~H.}\ \bibnamefont {Yuan}}, \bibinfo {author} {\bibfnamefont {Q.~L.}\ \bibnamefont {Dong}}, \bibinfo {author} {\bibfnamefont {Z.~M.}\ \bibnamefont {Sheng}}, \bibinfo {author} {\bibfnamefont {S.~V.}\ \bibnamefont {Bulanov}}, \bibinfo {author} {\bibfnamefont {Y.}~\bibnamefont {Kato}}, \bibinfo {author} {\bibfnamefont {J.}~\bibnamefont {Zhang}},\ and\ \bibinfo {author} {\bibfnamefont {T.}~\bibnamefont {Tajima}},\ }\bibfield  {title} {\bibinfo {title} {Study of x-ray emission enhancement via a high-contrast femtosecond laser interacting
  with a solid foil},\ }\href {https://doi.org/10.1103/PhysRevLett.100.045004} {\bibfield  {journal} {\bibinfo  {journal} {Phys. Rev. Lett.}\ }\textbf {\bibinfo {volume} {100}},\ \bibinfo {pages} {045004} (\bibinfo {year} {2008})}\BibitemShut {NoStop}%
\bibitem [{\citenamefont {Saemann}\ \emph {et~al.}(1999)\citenamefont {Saemann}, \citenamefont {Eidmann}, \citenamefont {Golovkin}, \citenamefont {Mancini}, \citenamefont {Andersson}, \citenamefont {F\"orster},\ and\ \citenamefont {Witte}}]{Saemann_PRL1999}%
  \BibitemOpen
  \bibfield  {author} {\bibinfo {author} {\bibfnamefont {A.}~\bibnamefont {Saemann}}, \bibinfo {author} {\bibfnamefont {K.}~\bibnamefont {Eidmann}}, \bibinfo {author} {\bibfnamefont {I.~E.}\ \bibnamefont {Golovkin}}, \bibinfo {author} {\bibfnamefont {R.~C.}\ \bibnamefont {Mancini}}, \bibinfo {author} {\bibfnamefont {E.}~\bibnamefont {Andersson}}, \bibinfo {author} {\bibfnamefont {E.}~\bibnamefont {F\"orster}},\ and\ \bibinfo {author} {\bibfnamefont {K.}~\bibnamefont {Witte}},\ }\bibfield  {title} {\bibinfo {title} {Isochoric heating of solid aluminum by ultrashort laser pulses focused on a tamped target},\ }\href {https://doi.org/10.1103/PhysRevLett.82.4843} {\bibfield  {journal} {\bibinfo  {journal} {Phys. Rev. Lett.}\ }\textbf {\bibinfo {volume} {82}},\ \bibinfo {pages} {4843} (\bibinfo {year} {1999})}\BibitemShut {NoStop}%
\bibitem [{\citenamefont {Hollinger}\ \emph {et~al.}(2020)\citenamefont {Hollinger}, \citenamefont {Wang}, \citenamefont {Wang}, \citenamefont {Moreau}, \citenamefont {Capeluto}, \citenamefont {Song}, \citenamefont {Rockwood}, \citenamefont {Bayarsaikhan}, \citenamefont {Kaymak}, \citenamefont {Pukhov}, \citenamefont {Shlyaptsev},\ and\ \citenamefont {Rocca}}]{Hollinger2020}%
  \BibitemOpen
  \bibfield  {author} {\bibinfo {author} {\bibfnamefont {R.}~\bibnamefont {Hollinger}}, \bibinfo {author} {\bibfnamefont {S.}~\bibnamefont {Wang}}, \bibinfo {author} {\bibfnamefont {Y.}~\bibnamefont {Wang}}, \bibinfo {author} {\bibfnamefont {A.}~\bibnamefont {Moreau}}, \bibinfo {author} {\bibfnamefont {M.~G.}\ \bibnamefont {Capeluto}}, \bibinfo {author} {\bibfnamefont {H.}~\bibnamefont {Song}}, \bibinfo {author} {\bibfnamefont {A.}~\bibnamefont {Rockwood}}, \bibinfo {author} {\bibfnamefont {E.}~\bibnamefont {Bayarsaikhan}}, \bibinfo {author} {\bibfnamefont {V.}~\bibnamefont {Kaymak}}, \bibinfo {author} {\bibfnamefont {A.}~\bibnamefont {Pukhov}}, \bibinfo {author} {\bibfnamefont {V.~N.}\ \bibnamefont {Shlyaptsev}},\ and\ \bibinfo {author} {\bibfnamefont {J.~J.}\ \bibnamefont {Rocca}},\ }\bibfield  {title} {\bibinfo {title} {Extreme ionization of heavy atoms in solid-density plasmas by relativistic second-harmonic laser pulses},\ }\href {https://doi.org/10.1038/s41566-020-0666-1} {\bibfield  {journal} {\bibinfo
   {journal} {Nature Photonics}\ }\textbf {\bibinfo {volume} {14}},\ \bibinfo {pages} {607} (\bibinfo {year} {2020})}\BibitemShut {NoStop}%
\bibitem [{\citenamefont {Kim}\ \emph {et~al.}(2005)\citenamefont {Kim}, \citenamefont {Kim}, \citenamefont {Kim}, \citenamefont {Lee}, \citenamefont {Lee}, \citenamefont {Park}, \citenamefont {Cho},\ and\ \citenamefont {Nam}}]{Kim_Jong_PRL2005_HHG_twocolor}%
  \BibitemOpen
  \bibfield  {author} {\bibinfo {author} {\bibfnamefont {I.~J.}\ \bibnamefont {Kim}}, \bibinfo {author} {\bibfnamefont {C.~M.}\ \bibnamefont {Kim}}, \bibinfo {author} {\bibfnamefont {H.~T.}\ \bibnamefont {Kim}}, \bibinfo {author} {\bibfnamefont {G.~H.}\ \bibnamefont {Lee}}, \bibinfo {author} {\bibfnamefont {Y.~S.}\ \bibnamefont {Lee}}, \bibinfo {author} {\bibfnamefont {J.~Y.}\ \bibnamefont {Park}}, \bibinfo {author} {\bibfnamefont {D.~J.}\ \bibnamefont {Cho}},\ and\ \bibinfo {author} {\bibfnamefont {C.~H.}\ \bibnamefont {Nam}},\ }\bibfield  {title} {\bibinfo {title} {Highly efficient high-harmonic generation in an orthogonally polarized two-color laser field},\ }\href {https://doi.org/10.1103/PhysRevLett.94.243901} {\bibfield  {journal} {\bibinfo  {journal} {Phys. Rev. Lett.}\ }\textbf {\bibinfo {volume} {94}},\ \bibinfo {pages} {243901} (\bibinfo {year} {2005})}\BibitemShut {NoStop}%
\bibitem [{\citenamefont {Yeung}\ \emph {et~al.}(2017)\citenamefont {Yeung}, \citenamefont {Rykovanov}, \citenamefont {Bierbach}, \citenamefont {Li}, \citenamefont {Eckner}, \citenamefont {Kuschel}, \citenamefont {Woldegeorgis}, \citenamefont {R{\"o}del}, \citenamefont {S{\"a}vert}, \citenamefont {Paulus}, \citenamefont {Coughlan}, \citenamefont {Dromey},\ and\ \citenamefont {Zepf}}]{Yeung_NatPhotonics_2017_attosecond_twocolor}%
  \BibitemOpen
  \bibfield  {author} {\bibinfo {author} {\bibfnamefont {M.}~\bibnamefont {Yeung}}, \bibinfo {author} {\bibfnamefont {S.}~\bibnamefont {Rykovanov}}, \bibinfo {author} {\bibfnamefont {J.}~\bibnamefont {Bierbach}}, \bibinfo {author} {\bibfnamefont {L.}~\bibnamefont {Li}}, \bibinfo {author} {\bibfnamefont {E.}~\bibnamefont {Eckner}}, \bibinfo {author} {\bibfnamefont {S.}~\bibnamefont {Kuschel}}, \bibinfo {author} {\bibfnamefont {A.}~\bibnamefont {Woldegeorgis}}, \bibinfo {author} {\bibfnamefont {C.}~\bibnamefont {R{\"o}del}}, \bibinfo {author} {\bibfnamefont {A.}~\bibnamefont {S{\"a}vert}}, \bibinfo {author} {\bibfnamefont {G.~G.}\ \bibnamefont {Paulus}}, \bibinfo {author} {\bibfnamefont {M.}~\bibnamefont {Coughlan}}, \bibinfo {author} {\bibfnamefont {B.}~\bibnamefont {Dromey}},\ and\ \bibinfo {author} {\bibfnamefont {M.}~\bibnamefont {Zepf}},\ }\bibfield  {title} {\bibinfo {title} {Experimental observation of attosecond control over relativistic electron bunches with two-colour fields},\ }\href
  {https://doi.org/10.1038/nphoton.2016.239} {\bibfield  {journal} {\bibinfo  {journal} {Nature Photonics}\ }\textbf {\bibinfo {volume} {11}},\ \bibinfo {pages} {32} (\bibinfo {year} {2017})}\BibitemShut {NoStop}%
\bibitem [{\citenamefont {Chen}\ \emph {et~al.}(2019)\citenamefont {Chen}, \citenamefont {He}, \citenamefont {Shaisultanov}, \citenamefont {Hatsagortsyan},\ and\ \citenamefont {Keitel}}]{Yue-Yue_PRL2019_positron_twocolor}%
  \BibitemOpen
  \bibfield  {author} {\bibinfo {author} {\bibfnamefont {Y.-Y.}\ \bibnamefont {Chen}}, \bibinfo {author} {\bibfnamefont {P.-L.}\ \bibnamefont {He}}, \bibinfo {author} {\bibfnamefont {R.}~\bibnamefont {Shaisultanov}}, \bibinfo {author} {\bibfnamefont {K.~Z.}\ \bibnamefont {Hatsagortsyan}},\ and\ \bibinfo {author} {\bibfnamefont {C.~H.}\ \bibnamefont {Keitel}},\ }\bibfield  {title} {\bibinfo {title} {Polarized positron beams via intense two-color laser pulses},\ }\href {https://doi.org/10.1103/PhysRevLett.123.174801} {\bibfield  {journal} {\bibinfo  {journal} {Phys. Rev. Lett.}\ }\textbf {\bibinfo {volume} {123}},\ \bibinfo {pages} {174801} (\bibinfo {year} {2019})}\BibitemShut {NoStop}%
\bibitem [{\citenamefont {Li}\ \emph {et~al.}(2019)\citenamefont {Li}, \citenamefont {Li}, \citenamefont {Ain}, \citenamefont {Hur}, \citenamefont {Ting}, \citenamefont {Kulagin}, \citenamefont {Kamperidis},\ and\ \citenamefont {Hafz}}]{Song_Li_SciAdv2019_accelerator_twocolor}%
  \BibitemOpen
  \bibfield  {author} {\bibinfo {author} {\bibfnamefont {S.}~\bibnamefont {Li}}, \bibinfo {author} {\bibfnamefont {G.}~\bibnamefont {Li}}, \bibinfo {author} {\bibfnamefont {Q.}~\bibnamefont {Ain}}, \bibinfo {author} {\bibfnamefont {M.~S.}\ \bibnamefont {Hur}}, \bibinfo {author} {\bibfnamefont {A.~C.}\ \bibnamefont {Ting}}, \bibinfo {author} {\bibfnamefont {V.~V.}\ \bibnamefont {Kulagin}}, \bibinfo {author} {\bibfnamefont {C.}~\bibnamefont {Kamperidis}},\ and\ \bibinfo {author} {\bibfnamefont {N.~A.~M.}\ \bibnamefont {Hafz}},\ }\bibfield  {title} {\bibinfo {title} {A laser-plasma accelerator driven by two-color relativistic femtosecond laser pulses},\ }\href {https://doi.org/10.1126/sciadv.aav7940} {\bibfield  {journal} {\bibinfo  {journal} {Science Advances}\ }\textbf {\bibinfo {volume} {5}},\ \bibinfo {pages} {eaav7940} (\bibinfo {year} {2019})}\BibitemShut {NoStop}%
\bibitem [{\citenamefont {Barbosa}\ \emph {et~al.}(2023)\citenamefont {Barbosa}, \citenamefont {Palastro}, \citenamefont {Ramsey}, \citenamefont {Weichman},\ and\ \citenamefont {Vranic}}]{Barbosa_arvix2023_breitwheeler_twocolor}%
  \BibitemOpen
  \bibfield  {author} {\bibinfo {author} {\bibfnamefont {B.}~\bibnamefont {Barbosa}}, \bibinfo {author} {\bibfnamefont {J.~P.}\ \bibnamefont {Palastro}}, \bibinfo {author} {\bibfnamefont {D.}~\bibnamefont {Ramsey}}, \bibinfo {author} {\bibfnamefont {K.}~\bibnamefont {Weichman}},\ and\ \bibinfo {author} {\bibfnamefont {M.}~\bibnamefont {Vranic}},\ }\href@noop {} {\bibinfo {title} {Phase control of nonlinear breit-wheeler pair creation}} (\bibinfo {year} {2023}),\ \Eprint {https://arxiv.org/abs/2310.13840} {arXiv:2310.13840 [physics.plasm-ph]} \BibitemShut {NoStop}%
\bibitem [{\citenamefont {Aparajit}\ \emph {et~al.}(2023)\citenamefont {Aparajit}, \citenamefont {Dulat}, \citenamefont {Choudhary}, \citenamefont {Lad}, \citenamefont {Ved}, \citenamefont {Couairon},\ and\ \citenamefont {Kumar}}]{Aparajit_APL2023}%
  \BibitemOpen
  \bibfield  {author} {\bibinfo {author} {\bibfnamefont {C.}~\bibnamefont {Aparajit}}, \bibinfo {author} {\bibfnamefont {A.}~\bibnamefont {Dulat}}, \bibinfo {author} {\bibfnamefont {A.}~\bibnamefont {Choudhary}}, \bibinfo {author} {\bibfnamefont {A.~D.}\ \bibnamefont {Lad}}, \bibinfo {author} {\bibfnamefont {Y.~M.}\ \bibnamefont {Ved}}, \bibinfo {author} {\bibfnamefont {A.}~\bibnamefont {Couairon}},\ and\ \bibinfo {author} {\bibfnamefont {G.~R.}\ \bibnamefont {Kumar}},\ }\bibfield  {title} {\bibinfo {title} {{The femtosecond structure of extreme contrast, multi-terawatt second-harmonic laser pulses at 400 nm}},\ }\href {https://doi.org/10.1063/5.0174513} {\bibfield  {journal} {\bibinfo  {journal} {Applied Physics Letters}\ }\textbf {\bibinfo {volume} {123}},\ \bibinfo {pages} {141108} (\bibinfo {year} {2023})}\BibitemShut {NoStop}%
\bibitem [{\citenamefont {Aparajit}\ \emph {et~al.}(2021)\citenamefont {Aparajit}, \citenamefont {Jana}, \citenamefont {Lad}, \citenamefont {Ved}, \citenamefont {Couairon},\ and\ \citenamefont {Kumar}}]{Aparajit:21}%
  \BibitemOpen
  \bibfield  {author} {\bibinfo {author} {\bibfnamefont {C.}~\bibnamefont {Aparajit}}, \bibinfo {author} {\bibfnamefont {K.}~\bibnamefont {Jana}}, \bibinfo {author} {\bibfnamefont {A.~D.}\ \bibnamefont {Lad}}, \bibinfo {author} {\bibfnamefont {Y.~M.}\ \bibnamefont {Ved}}, \bibinfo {author} {\bibfnamefont {A.}~\bibnamefont {Couairon}},\ and\ \bibinfo {author} {\bibfnamefont {G.~R.}\ \bibnamefont {Kumar}},\ }\bibfield  {title} {\bibinfo {title} {Efficient second-harmonic generation of a high-energy, femtosecond laser pulse in a lithium triborate crystal},\ }\href {https://doi.org/10.1364/OL.423725} {\bibfield  {journal} {\bibinfo  {journal} {Opt. Lett.}\ }\textbf {\bibinfo {volume} {46}},\ \bibinfo {pages} {3540} (\bibinfo {year} {2021})}\BibitemShut {NoStop}%
\bibitem [{\citenamefont {Hurricane}\ \emph {et~al.}(2014)\citenamefont {Hurricane}, \citenamefont {Callahan}, \citenamefont {Casey}, \citenamefont {Celliers}, \citenamefont {Cerjan}, \citenamefont {Dewald}, \citenamefont {Dittrich}, \citenamefont {D{\"o}ppner}, \citenamefont {Hinkel}, \citenamefont {Hopkins}, \citenamefont {Kline}, \citenamefont {Le~Pape}, \citenamefont {Ma}, \citenamefont {MacPhee}, \citenamefont {Milovich}, \citenamefont {Pak}, \citenamefont {Park}, \citenamefont {Patel}, \citenamefont {Remington}, \citenamefont {Salmonson}, \citenamefont {Springer},\ and\ \citenamefont {Tommasini}}]{Hurricane2014}%
  \BibitemOpen
  \bibfield  {author} {\bibinfo {author} {\bibfnamefont {O.~A.}\ \bibnamefont {Hurricane}}, \bibinfo {author} {\bibfnamefont {D.~A.}\ \bibnamefont {Callahan}}, \bibinfo {author} {\bibfnamefont {D.~T.}\ \bibnamefont {Casey}}, \bibinfo {author} {\bibfnamefont {P.~M.}\ \bibnamefont {Celliers}}, \bibinfo {author} {\bibfnamefont {C.}~\bibnamefont {Cerjan}}, \bibinfo {author} {\bibfnamefont {E.~L.}\ \bibnamefont {Dewald}}, \bibinfo {author} {\bibfnamefont {T.~R.}\ \bibnamefont {Dittrich}}, \bibinfo {author} {\bibfnamefont {T.}~\bibnamefont {D{\"o}ppner}}, \bibinfo {author} {\bibfnamefont {D.~E.}\ \bibnamefont {Hinkel}}, \bibinfo {author} {\bibfnamefont {L.~F.~B.}\ \bibnamefont {Hopkins}}, \bibinfo {author} {\bibfnamefont {J.~L.}\ \bibnamefont {Kline}}, \bibinfo {author} {\bibfnamefont {S.}~\bibnamefont {Le~Pape}}, \bibinfo {author} {\bibfnamefont {T.}~\bibnamefont {Ma}}, \bibinfo {author} {\bibfnamefont {A.~G.}\ \bibnamefont {MacPhee}}, \bibinfo {author} {\bibfnamefont {J.~L.}\ \bibnamefont {Milovich}}, \bibinfo
  {author} {\bibfnamefont {A.}~\bibnamefont {Pak}}, \bibinfo {author} {\bibfnamefont {H.-S.}\ \bibnamefont {Park}}, \bibinfo {author} {\bibfnamefont {P.~K.}\ \bibnamefont {Patel}}, \bibinfo {author} {\bibfnamefont {B.~A.}\ \bibnamefont {Remington}}, \bibinfo {author} {\bibfnamefont {J.~D.}\ \bibnamefont {Salmonson}}, \bibinfo {author} {\bibfnamefont {P.~T.}\ \bibnamefont {Springer}},\ and\ \bibinfo {author} {\bibfnamefont {R.}~\bibnamefont {Tommasini}},\ }\bibfield  {title} {\bibinfo {title} {Fuel gain exceeding unity in an inertially confined fusion implosion},\ }\href {https://doi.org/10.1038/nature13008} {\bibfield  {journal} {\bibinfo  {journal} {Nature}\ }\textbf {\bibinfo {volume} {506}},\ \bibinfo {pages} {343} (\bibinfo {year} {2014})}\BibitemShut {NoStop}%
\bibitem [{\citenamefont {Cojocaru}\ \emph {et~al.}(2016)\citenamefont {Cojocaru}, \citenamefont {Ungureanu}, \citenamefont {Banici}, \citenamefont {Ursescu}, \citenamefont {Guilbaud}, \citenamefont {Delmas}, \citenamefont {Marec}, \citenamefont {Neveu}, \citenamefont {Demailly}, \citenamefont {Pittman}, \citenamefont {Kazamias}, \citenamefont {Daboussi}, \citenamefont {Cassou}, \citenamefont {Li}, \citenamefont {Klisnick}, \citenamefont {Zeitoun},\ and\ \citenamefont {Ros}}]{Cojocaru:16}%
  \BibitemOpen
  \bibfield  {author} {\bibinfo {author} {\bibfnamefont {G.~V.}\ \bibnamefont {Cojocaru}}, \bibinfo {author} {\bibfnamefont {R.~G.}\ \bibnamefont {Ungureanu}}, \bibinfo {author} {\bibfnamefont {R.~A.}\ \bibnamefont {Banici}}, \bibinfo {author} {\bibfnamefont {D.}~\bibnamefont {Ursescu}}, \bibinfo {author} {\bibfnamefont {O.}~\bibnamefont {Guilbaud}}, \bibinfo {author} {\bibfnamefont {O.}~\bibnamefont {Delmas}}, \bibinfo {author} {\bibfnamefont {A.~L.}\ \bibnamefont {Marec}}, \bibinfo {author} {\bibfnamefont {O.}~\bibnamefont {Neveu}}, \bibinfo {author} {\bibfnamefont {J.}~\bibnamefont {Demailly}}, \bibinfo {author} {\bibfnamefont {M.}~\bibnamefont {Pittman}}, \bibinfo {author} {\bibfnamefont {S.}~\bibnamefont {Kazamias}}, \bibinfo {author} {\bibfnamefont {S.}~\bibnamefont {Daboussi}}, \bibinfo {author} {\bibfnamefont {K.}~\bibnamefont {Cassou}}, \bibinfo {author} {\bibfnamefont {L.}~\bibnamefont {Li}}, \bibinfo {author} {\bibfnamefont {A.}~\bibnamefont {Klisnick}}, \bibinfo {author} {\bibfnamefont
  {P.}~\bibnamefont {Zeitoun}},\ and\ \bibinfo {author} {\bibfnamefont {D.}~\bibnamefont {Ros}},\ }\bibfield  {title} {\bibinfo {title} {One long and two short pumping pulses control for plasma x-ray amplifier optimization},\ }\href {https://doi.org/10.1364/OE.24.014260} {\bibfield  {journal} {\bibinfo  {journal} {Opt. Express}\ }\textbf {\bibinfo {volume} {24}},\ \bibinfo {pages} {14260} (\bibinfo {year} {2016})}\BibitemShut {NoStop}%
\bibitem [{\citenamefont {Nowak}\ \emph {et~al.}(2016)\citenamefont {Nowak}, \citenamefont {Kurosawa}, \citenamefont {Suganuma}, \citenamefont {Kawasuji}, \citenamefont {Nakarai}, \citenamefont {Saito}, \citenamefont {Fujimoto},\ and\ \citenamefont {Mizoguchi}}]{Nowak:16}%
  \BibitemOpen
  \bibfield  {author} {\bibinfo {author} {\bibfnamefont {K.~M.}\ \bibnamefont {Nowak}}, \bibinfo {author} {\bibfnamefont {Y.}~\bibnamefont {Kurosawa}}, \bibinfo {author} {\bibfnamefont {T.}~\bibnamefont {Suganuma}}, \bibinfo {author} {\bibfnamefont {Y.}~\bibnamefont {Kawasuji}}, \bibinfo {author} {\bibfnamefont {H.}~\bibnamefont {Nakarai}}, \bibinfo {author} {\bibfnamefont {T.}~\bibnamefont {Saito}}, \bibinfo {author} {\bibfnamefont {J.}~\bibnamefont {Fujimoto}},\ and\ \bibinfo {author} {\bibfnamefont {H.}~\bibnamefont {Mizoguchi}},\ }\bibfield  {title} {\bibinfo {title} {Synthesis of arbitrary pulse waveforms in qcl-seeded ns-pulse co2 laser for optimization of an lpp euv source},\ }\href {https://doi.org/10.1364/OL.41.003118} {\bibfield  {journal} {\bibinfo  {journal} {Opt. Lett.}\ }\textbf {\bibinfo {volume} {41}},\ \bibinfo {pages} {3118} (\bibinfo {year} {2016})}\BibitemShut {NoStop}%
\bibitem [{\citenamefont {Meijer}\ \emph {et~al.}(2017)\citenamefont {Meijer}, \citenamefont {Stodolna}, \citenamefont {Eikema},\ and\ \citenamefont {Witte}}]{Meijer:17}%
  \BibitemOpen
  \bibfield  {author} {\bibinfo {author} {\bibfnamefont {R.~A.}\ \bibnamefont {Meijer}}, \bibinfo {author} {\bibfnamefont {A.~S.}\ \bibnamefont {Stodolna}}, \bibinfo {author} {\bibfnamefont {K.~S.~E.}\ \bibnamefont {Eikema}},\ and\ \bibinfo {author} {\bibfnamefont {S.}~\bibnamefont {Witte}},\ }\bibfield  {title} {\bibinfo {title} {High-energy nd:yag laser system with arbitrary sub-nanosecond pulse shaping capability},\ }\href {https://doi.org/10.1364/OL.42.002758} {\bibfield  {journal} {\bibinfo  {journal} {Opt. Lett.}\ }\textbf {\bibinfo {volume} {42}},\ \bibinfo {pages} {2758} (\bibinfo {year} {2017})}\BibitemShut {NoStop}%
\bibitem [{\citenamefont {Pangovski}\ \emph {et~al.}(2014)\citenamefont {Pangovski}, \citenamefont {Sparkes}, \citenamefont {Cockburn}, \citenamefont {O’Neill}, \citenamefont {Teh}, \citenamefont {Lin},\ and\ \citenamefont {Richardson}}]{Pangovski_2014}%
  \BibitemOpen
  \bibfield  {author} {\bibinfo {author} {\bibfnamefont {K.}~\bibnamefont {Pangovski}}, \bibinfo {author} {\bibfnamefont {M.}~\bibnamefont {Sparkes}}, \bibinfo {author} {\bibfnamefont {A.}~\bibnamefont {Cockburn}}, \bibinfo {author} {\bibfnamefont {W.}~\bibnamefont {O’Neill}}, \bibinfo {author} {\bibfnamefont {P.~S.}\ \bibnamefont {Teh}}, \bibinfo {author} {\bibfnamefont {D.}~\bibnamefont {Lin}},\ and\ \bibinfo {author} {\bibfnamefont {D.}~\bibnamefont {Richardson}},\ }\bibfield  {title} {\bibinfo {title} {Control of material transport through pulse shape manipulation—a development toward designer pulses},\ }\href {https://doi.org/10.1109/JSTQE.2014.2302441} {\bibfield  {journal} {\bibinfo  {journal} {IEEE Journal of Selected Topics in Quantum Electronics}\ }\textbf {\bibinfo {volume} {20}},\ \bibinfo {pages} {51} (\bibinfo {year} {2014})}\BibitemShut {NoStop}%
\bibitem [{\citenamefont {Warren}\ \emph {et~al.}(1993)\citenamefont {Warren}, \citenamefont {Rabitz},\ and\ \citenamefont {Dahleh}}]{Warren_Science1993}%
  \BibitemOpen
  \bibfield  {author} {\bibinfo {author} {\bibfnamefont {W.~S.}\ \bibnamefont {Warren}}, \bibinfo {author} {\bibfnamefont {H.}~\bibnamefont {Rabitz}},\ and\ \bibinfo {author} {\bibfnamefont {M.}~\bibnamefont {Dahleh}},\ }\bibfield  {title} {\bibinfo {title} {Coherent control of quantum dynamics: The dream is alive},\ }\href {https://doi.org/10.1126/science.259.5101.1581} {\bibfield  {journal} {\bibinfo  {journal} {Science}\ }\textbf {\bibinfo {volume} {259}},\ \bibinfo {pages} {1581} (\bibinfo {year} {1993})}\BibitemShut {NoStop}%
\bibitem [{\citenamefont {Assion}\ \emph {et~al.}(1998)\citenamefont {Assion}, \citenamefont {Baumert}, \citenamefont {Bergt}, \citenamefont {Brixner}, \citenamefont {Kiefer}, \citenamefont {Seyfried}, \citenamefont {Strehle},\ and\ \citenamefont {Gerber}}]{Assion_Science1998}%
  \BibitemOpen
  \bibfield  {author} {\bibinfo {author} {\bibfnamefont {A.}~\bibnamefont {Assion}}, \bibinfo {author} {\bibfnamefont {T.}~\bibnamefont {Baumert}}, \bibinfo {author} {\bibfnamefont {M.}~\bibnamefont {Bergt}}, \bibinfo {author} {\bibfnamefont {T.}~\bibnamefont {Brixner}}, \bibinfo {author} {\bibfnamefont {B.}~\bibnamefont {Kiefer}}, \bibinfo {author} {\bibfnamefont {V.}~\bibnamefont {Seyfried}}, \bibinfo {author} {\bibfnamefont {M.}~\bibnamefont {Strehle}},\ and\ \bibinfo {author} {\bibfnamefont {G.}~\bibnamefont {Gerber}},\ }\bibfield  {title} {\bibinfo {title} {Control of chemical reactions by feedback-optimized phase-shaped femtosecond laser pulses},\ }\href {https://doi.org/10.1126/science.282.5390.919} {\bibfield  {journal} {\bibinfo  {journal} {Science}\ }\textbf {\bibinfo {volume} {282}},\ \bibinfo {pages} {919} (\bibinfo {year} {1998})}\BibitemShut {NoStop}%
\bibitem [{\citenamefont {Riconda}\ \emph {et~al.}(2015)\citenamefont {Riconda}, \citenamefont {Raynaud}, \citenamefont {Vialis},\ and\ \citenamefont {Grech}}]{Riconda_PoP2015_SPW_electron_scaling}%
  \BibitemOpen
  \bibfield  {author} {\bibinfo {author} {\bibfnamefont {C.}~\bibnamefont {Riconda}}, \bibinfo {author} {\bibfnamefont {M.}~\bibnamefont {Raynaud}}, \bibinfo {author} {\bibfnamefont {T.}~\bibnamefont {Vialis}},\ and\ \bibinfo {author} {\bibfnamefont {M.}~\bibnamefont {Grech}},\ }\bibfield  {title} {\bibinfo {title} {{Simple scalings for various regimes of electron acceleration in surface plasma waves}},\ }\href {https://doi.org/10.1063/1.4923443} {\bibfield  {journal} {\bibinfo  {journal} {Physics of Plasmas}\ }\textbf {\bibinfo {volume} {22}},\ \bibinfo {pages} {073103} (\bibinfo {year} {2015})}\BibitemShut {NoStop}%
\bibitem [{\citenamefont {Raynaud}\ \emph {et~al.}(2007)\citenamefont {Raynaud}, \citenamefont {Kupersztych}, \citenamefont {Riconda}, \citenamefont {Adam},\ and\ \citenamefont {Héron}}]{Raynaud_PoP2007_SPW}%
  \BibitemOpen
  \bibfield  {author} {\bibinfo {author} {\bibfnamefont {M.}~\bibnamefont {Raynaud}}, \bibinfo {author} {\bibfnamefont {J.}~\bibnamefont {Kupersztych}}, \bibinfo {author} {\bibfnamefont {C.}~\bibnamefont {Riconda}}, \bibinfo {author} {\bibfnamefont {J.~C.}\ \bibnamefont {Adam}},\ and\ \bibinfo {author} {\bibfnamefont {A.}~\bibnamefont {Héron}},\ }\bibfield  {title} {\bibinfo {title} {{Strongly enhanced laser absorption and electron acceleration via resonant excitation of surface plasma waves}},\ }\href {https://doi.org/10.1063/1.2755969} {\bibfield  {journal} {\bibinfo  {journal} {Physics of Plasmas}\ }\textbf {\bibinfo {volume} {14}},\ \bibinfo {pages} {092702} (\bibinfo {year} {2007})}\BibitemShut {NoStop}%
\bibitem [{\citenamefont {Kupersztych}\ \emph {et~al.}(2004)\citenamefont {Kupersztych}, \citenamefont {Raynaud},\ and\ \citenamefont {Riconda}}]{Kupersztych_Raynaud_Riconda_PoP2004_SPW}%
  \BibitemOpen
  \bibfield  {author} {\bibinfo {author} {\bibfnamefont {J.}~\bibnamefont {Kupersztych}}, \bibinfo {author} {\bibfnamefont {M.}~\bibnamefont {Raynaud}},\ and\ \bibinfo {author} {\bibfnamefont {C.}~\bibnamefont {Riconda}},\ }\bibfield  {title} {\bibinfo {title} {{Electron acceleration by surface plasma waves in the interaction between femtosecond laser pulses and sharp-edged overdense plasmas}},\ }\href {https://doi.org/10.1063/1.1650353} {\bibfield  {journal} {\bibinfo  {journal} {Physics of Plasmas}\ }\textbf {\bibinfo {volume} {11}},\ \bibinfo {pages} {1669} (\bibinfo {year} {2004})}\BibitemShut {NoStop}%
\bibitem [{\citenamefont {Raynaud}\ \emph {et~al.}(2020)\citenamefont {Raynaud}, \citenamefont {H{\'e}ron},\ and\ \citenamefont {Adam}}]{Raynaud2020}%
  \BibitemOpen
  \bibfield  {author} {\bibinfo {author} {\bibfnamefont {M.}~\bibnamefont {Raynaud}}, \bibinfo {author} {\bibfnamefont {A.}~\bibnamefont {H{\'e}ron}},\ and\ \bibinfo {author} {\bibfnamefont {J.-C.}\ \bibnamefont {Adam}},\ }\bibfield  {title} {\bibinfo {title} {Excitation of surface plasma waves and fast electron generation in relativistic laser--plasma interaction},\ }\href {https://doi.org/10.1038/s41598-020-70221-9} {\bibfield  {journal} {\bibinfo  {journal} {Scientific Reports}\ }\textbf {\bibinfo {volume} {10}},\ \bibinfo {pages} {13450} (\bibinfo {year} {2020})}\BibitemShut {NoStop}%
\bibitem [{\citenamefont {Bigongiari}\ \emph {et~al.}(2013)\citenamefont {Bigongiari}, \citenamefont {Raynaud}, \citenamefont {Riconda},\ and\ \citenamefont {Héron}}]{Bigongiari_PoP2013}%
  \BibitemOpen
  \bibfield  {author} {\bibinfo {author} {\bibfnamefont {A.}~\bibnamefont {Bigongiari}}, \bibinfo {author} {\bibfnamefont {M.}~\bibnamefont {Raynaud}}, \bibinfo {author} {\bibfnamefont {C.}~\bibnamefont {Riconda}},\ and\ \bibinfo {author} {\bibfnamefont {A.}~\bibnamefont {Héron}},\ }\bibfield  {title} {\bibinfo {title} {{Improved ion acceleration via laser surface plasma waves excitation}},\ }\href {https://doi.org/10.1063/1.4802989} {\bibfield  {journal} {\bibinfo  {journal} {Physics of Plasmas}\ }\textbf {\bibinfo {volume} {20}},\ \bibinfo {pages} {052701} (\bibinfo {year} {2013})}\BibitemShut {NoStop}%
\bibitem [{\citenamefont {Marini}\ \emph {et~al.}(2021)\citenamefont {Marini}, \citenamefont {Kleij}, \citenamefont {Amiranoff}, \citenamefont {Grech}, \citenamefont {Riconda},\ and\ \citenamefont {Raynaud}}]{Marini_PoP2021}%
  \BibitemOpen
  \bibfield  {author} {\bibinfo {author} {\bibfnamefont {S.}~\bibnamefont {Marini}}, \bibinfo {author} {\bibfnamefont {P.~S.}\ \bibnamefont {Kleij}}, \bibinfo {author} {\bibfnamefont {F.}~\bibnamefont {Amiranoff}}, \bibinfo {author} {\bibfnamefont {M.}~\bibnamefont {Grech}}, \bibinfo {author} {\bibfnamefont {C.}~\bibnamefont {Riconda}},\ and\ \bibinfo {author} {\bibfnamefont {M.}~\bibnamefont {Raynaud}},\ }\bibfield  {title} {\bibinfo {title} {{Key parameters for surface plasma wave excitation in the ultra-high intensity regime}},\ }\href {https://doi.org/10.1063/5.0052599} {\bibfield  {journal} {\bibinfo  {journal} {Physics of Plasmas}\ }\textbf {\bibinfo {volume} {28}},\ \bibinfo {pages} {073104} (\bibinfo {year} {2021})}\BibitemShut {NoStop}%
\bibitem [{\citenamefont {Derouillat}\ \emph {et~al.}(2018)\citenamefont {Derouillat}, \citenamefont {Beck}, \citenamefont {Pérez}, \citenamefont {Vinci}, \citenamefont {Chiaramello}, \citenamefont {Grassi}, \citenamefont {Flé}, \citenamefont {Bouchard}, \citenamefont {Plotnikov}, \citenamefont {Aunai}, \citenamefont {Dargent}, \citenamefont {Riconda},\ and\ \citenamefont {Grech}}]{Smilei_DEROUILLAT2018351}%
  \BibitemOpen
  \bibfield  {author} {\bibinfo {author} {\bibfnamefont {J.}~\bibnamefont {Derouillat}}, \bibinfo {author} {\bibfnamefont {A.}~\bibnamefont {Beck}}, \bibinfo {author} {\bibfnamefont {F.}~\bibnamefont {Pérez}}, \bibinfo {author} {\bibfnamefont {T.}~\bibnamefont {Vinci}}, \bibinfo {author} {\bibfnamefont {M.}~\bibnamefont {Chiaramello}}, \bibinfo {author} {\bibfnamefont {A.}~\bibnamefont {Grassi}}, \bibinfo {author} {\bibfnamefont {M.}~\bibnamefont {Flé}}, \bibinfo {author} {\bibfnamefont {G.}~\bibnamefont {Bouchard}}, \bibinfo {author} {\bibfnamefont {I.}~\bibnamefont {Plotnikov}}, \bibinfo {author} {\bibfnamefont {N.}~\bibnamefont {Aunai}}, \bibinfo {author} {\bibfnamefont {J.}~\bibnamefont {Dargent}}, \bibinfo {author} {\bibfnamefont {C.}~\bibnamefont {Riconda}},\ and\ \bibinfo {author} {\bibfnamefont {M.}~\bibnamefont {Grech}},\ }\bibfield  {title} {\bibinfo {title} {Smilei : A collaborative, open-source, multi-purpose particle-in-cell code for plasma simulation},\ }\href
  {https://doi.org/https://doi.org/10.1016/j.cpc.2017.09.024} {\bibfield  {journal} {\bibinfo  {journal} {Computer Physics Communications}\ }\textbf {\bibinfo {volume} {222}},\ \bibinfo {pages} {351} (\bibinfo {year} {2018})}\BibitemShut {NoStop}%
\bibitem [{\citenamefont {Rajeev}\ \emph {et~al.}(2004)\citenamefont {Rajeev}, \citenamefont {Ayyub}, \citenamefont {Bagchi},\ and\ \citenamefont {Kumar}}]{Rajeev_OL2004}%
  \BibitemOpen
  \bibfield  {author} {\bibinfo {author} {\bibfnamefont {P.~P.}\ \bibnamefont {Rajeev}}, \bibinfo {author} {\bibfnamefont {P.}~\bibnamefont {Ayyub}}, \bibinfo {author} {\bibfnamefont {S.}~\bibnamefont {Bagchi}},\ and\ \bibinfo {author} {\bibfnamefont {G.~R.}\ \bibnamefont {Kumar}},\ }\bibfield  {title} {\bibinfo {title} {Nanostructures, local fields, and enhanced absorption in intense light--matter interaction},\ }\href {https://doi.org/10.1364/OL.29.002662} {\bibfield  {journal} {\bibinfo  {journal} {Opt. Lett.}\ }\textbf {\bibinfo {volume} {29}},\ \bibinfo {pages} {2662} (\bibinfo {year} {2004})}\BibitemShut {NoStop}%
\end{thebibliography}%

\end{document}